\documentclass{acmrip}
\usepackage[boxed, lined]{algorithm2e}
\usepackage[tight,footnotesize]{subfigure}

    \usepackage[dvips]{epsfig}

    \usepackage[latin1]{inputenc}

    \newdef{definition}[theorem]{Definition}
    \newdef{remark}[theorem]{Remark}

    \usepackage{color}
    
    \newcommand{\figref}[1]{Figure~\ref{fig:#1}}
    \newcommand{\tabref}[1]{Table~\ref{tab:#1}}

    \newcommand{\secref}[1]{Section~\ref{sec:#1}}

    \newcommand{\AutoCast}[0]{{\em AutoCast} }

    \title{Empowered by Wireless Communication:
Distributed Methods for Self-Organizing Traffic Collectives}
    \author{S\'{A}NDOR P. FEKETE  and CHRISTIANE SCHMIDT \\ Braunschweig University of Technology\and AXEL WEGENER, HORST HELLBR\"UCK and STEFAN FISCHER \\ University of L\"ubeck }
    \begin{abstract}
In recent years, tremendous progress has been made in understanding
the dynamics of vehicle traffic flow and traffic congestion by
interpreting traffic as a multi-particle system. This helps to
explain the onset and persistence of many undesired phenomena, e.g.,
traffic jams. It also reflects the apparent helplessness of drivers
in traffic, who feel like passive particles that are pushed around
by exterior forces; one of the crucial aspects is
the inability to communicate and coordinate with other traffic
participants.

We present distributed methods for solving these fundamental
problems, employing modern wireless, ad-hoc, multi-hop networks.
The underlying idea is to use these capabilities as the basis for
self-organizing methods for coordinating data collection and
processing, recognizing traffic phenomena, and changing their
structure by coordinated behavior. The overall objective is a
multi-level approach that reaches from protocols for
local wireless communication, data dissemination, pattern
recognition, over hierarchical structuring and coordinated behavior,
all the way to large-scale traffic regulation.

In this article we
describe three types of results: (i) self-organizing and distributed
methods for maintaining and collecting data (using our concept of
{\em Hovering Data Clouds}); (ii) adaptive data dissemination for
traffic information systems; (iii) methods for self-recognition of
traffic jams. We conclude by describing higher-level aspects of our work.

\end{abstract}
    
\begin{document}
    \begin{bottomstuff}
    The work described in this article is based in parts on the conference papers \cite{fsw+-hdcrtj-06} and \cite{We2007AutoCast}. It has been supported within the
DFG Priority Programme ``Organic
    Computing'' (SPP 1183), project ``AutoNomos'', grant numbers FE 407/11-1, FE 407/11-2, FE 407/11-3, Fi 605/12-1, Fi 605/12-2, Fi 605/12-3.\\
    Authors' addresses: S.\ P.\ Fekete and C.\ Schmidt, Department of Computer Science, Braunschweig
    University of Technology, Braunschweig, Germany,
    email:\{s.fekete,c.schmidt\}@tu-bs.de; A.\ Wegener, H.\ Hellbr\"uck and
    S.\ Fischer, Institute of Telematics, University of L\"ubeck,
    L\"ubeck, Germany, email:\{wegener,hellbrueck,fischer\}@itm.uni-luebeck.de.
    \end{bottomstuff}
    \maketitle
    %%%%%%%%%%%%%%%%%%%%%%%%%%%%%%%%%%%%%%%%%%%%%%%%%%%%%%%%%%%
    %introduction
    %%%%%%%%%%%%%%%%%%%%%%%%%%%%%%%%%%%%%%%%%%%%%%%%%%%%%%%%%%%
    \section{Introduction}

\subsection{Complex Systems and Organic Computing}
A standard scientific method for {\em understanding} complicated situations is
to analyze them in a top-down, hierarchical manner.  This also works
well for {\em organizing} a large variety of structures; that is why a similar
approach has worked extremely well for employing
computers in so many aspects of our life.

On the other hand, our world has grown to be increasingly complex. The
resulting challenges have become so demanding that it is impossible to ignore
that a large variety of systems has a very different structure: The stability
and effectiveness of our modern political, social and economic structures
relies on the fact that they are based on decentralized, distributed and
self-organizing mechanisms. (In this context, see~\cite{s-wc-04} for a recent
non-fiction bestseller.)

Until not very long ago, scientific efforts for studying computing methodologies
for decentralized complex systems have been very limited.  Traditional
computing systems are based on a centralized algorithmic paradigm: data is
gathered, processed, and the result is administered by one central authority.
Each of these aspects is subject to obstructions. On the other hand, ``Living
organisms [...] are to be treated as dynamic systems and contain all
infrastructure necessary for their development, instead of depending on
coupling to a separate thinking mind. We call this computing paradigm {\em
organic computing} to emphasize both organic structure and complex, purposeful
action'' \cite{org}.  The importance of organic computing has been motivated as
follows: ``The advantages of self-organizing systems are obvious: They behave
more like intelligent assistants, rather than a rigidly programmed slave. They
are flexible, robust against (partial) failure, and are able to self-optimize.
The cost of design decreases, because not every variant has to be programmed in
advance'' \cite{msu-ocpfp-04}.

Two important properties of organic computing systems are {\em
self-organization} and {\em emergence}. We follow the definition of de Wolf
\cite{wolf05} who gives the following definitions:
\begin{quote}
``Self-organization is a dynamical and adaptive process where systems acquire
and maintain structure themselves, without external control.''
\end{quote}
and
\begin{quote}
``A system exhibits emergence when there are coherent emergents at the
macro-level that dynamically arise from the interactions between the parts
at the micro-level. Such emergents are novel w.r.t. the individual parts of
the system.''
\end{quote}

This work was conducted in the spirit of organic computing. In the following we
will show how to combine aspects of distributed computing and communication (introduced in Section 1.2)
with a new concept of algorithms and data structures, in order to deal with the
challenge of influencing a very important complex system: traffic.  Clearly,
road traffic by itself (as introduced in Section 1.3) exhibits both properties postulated by de Wolf. Vehicles
interact on the micro-level without external control by local interactions
(self-organized) and due to this micro-level interactions structures at the
macro-level arise, such as traffic jams or convoys.

\subsection{Distributed Communication}
A critical aspect for influencing a complex system, no matter whether in a
centralized or decentralized manner, is making use of distributed
communication, which is becoming more and more feasible by the advance of
modern technology.  Today's applications for distributed systems rely on
unicast or multicast communication, where communication partners are identified
by layer-3 addresses. Prominent and well-known examples are
Client-Server-Applications like webbrow\-sing. When mobile ad-hoc networks
(MANETs) first came up, research concentrated on efficient routing to keep up
the existing communication paradigm. By increasing the number of nodes and
introducing dynamics in the network due to node mobility or switching nodes on
and off in a large multi-hop network, routing based on addresses became a hard
challenge. In the meantime systems have grown larger, to the point where they consist of thousands of cheap
battery-powered devices that organize themselves and open a new research field
called {\em sensor networks}. Limitations of bandwidth, computing power and storage
in sensor networks drive a new communication paradigm. While these systems need
to be designed for frequent node failures, redundancy was introduced in the
network, decreasing the importance of single nodes and addresses. The
stringent separation of the lower communication layers has also become open for discussion, as
it introduces computational and networking overhead especially when
identification of nodes loses importance.

\subsection{Traffic}
Traffic is one of the most influential phenomena of civilization. On
a small scale, it affects the daily life of billions of individuals;
on a larger scale, it determines the operating conditions of all
industrialized economies; and on the global scale, traffic has a
tremendous impact on the living conditions on our planet, both in a
positive and in a negative way.

All this makes traffic one of the most important complex systems of
our modern world. It has several levels of complexity, reaching from
individual actions of the drivers, over local phenomena like density
fluctuations and traffic jams, traffic participants' choice of
transport mode and time, regional and temporal traffic patterns, all
the way up to long-range traffic development and regulation.

In recent years, tremendous progress has been made in understanding
the dynamics of traffic flow and traffic congestion. Arguably the
most significant contribution has come from physics, interpreting
traffic as a mul\-ti-par\-ti\-cle system. These models explain how
the complexity of traffic emerges from the self-organization of
individuals that follow simple rules. They also meet the popular
appeal of systems of interacting, autonomous agents as
problem-solving devices and models of complex behavior. In short,
the ``decentralized'' view, which goes beyond attempts at
centralized simulation and control, has improved our understanding
of traffic.

\subsection{Combining Communication and Mobility}
Modern hardware has advanced to the point that it is technically
possible to enable communication and coordination between traffic
participants. At this point, this is mostly seen as a mere technical
gadget to facilitate driving, but the far-reaching, large-scale
consequences have yet to be explored. Making use of the technical
possibilities of communication and coordination should allow
significant changes in the large-scale behavior of traffic as a
complex network phenomenon. However, combining mobility and
communication for coordinated behavior does not only solve problems,
it also creates new ones, as it is a challenge in itself to maintain
the involved ad-hoc networks, as well as the related information
that is independent of individual vehicles.

To the best of our knowledge, the idea of combining the above
aspects, i.e., self-organizing traffic by combining ad-hoc networks
with distributed decentralized algorithms has not been studied so
far. This may be because it requires combining a number of different
aspects, each of which has only been developed by itself in recent
years: mobile ad-hoc networks, models for large systems of
self-driven multi-particle systems, as well as algorithmic aspects
of decentralized distributed computing, possibly with elements of
game-theoretic interaction.

\subsection{Our Work}
Our basic idea is to develop a
decentralized method for traffic analysis and control, based on a
bottom-up, multilevel approach. Beyond the motivations described
above, it should be stressed that an aspect of particular relevance
is {\em scalability:} while the computational effort for a
centralized approach increases prohibitively with the number of
vehicles, a decentralized method relies on neighborhood interaction
of constant size.

At this point, we focus on a number of different aspects:
\begin{enumerate}
\item One is the key concept of {\em
Hovering Data Clouds} (HDCs), which are virtual structures that
exist independent of particular carriers.
\item Data dissemination in large-scale ad-hoc networks formed by
moving vehicles in a complex traffic situation requires adaptive
protocols for traffic information systems.
\item The self-recognition of traffic situations by distributed reasoning,
and without a central operating center allows drivers to benefit
from joint identification of traffic jam types and parameters.
\item Based on these insights and structures, we can aim at actually
{\em changing} the behavior of traffic.
\end{enumerate}

The rest of this paper is organized as follows. In the next
Section~2, we give a broad survey of related work, subdivided into distributed communication
and data dissemination (2.1), traffic and telematics (2.2), and 
traffic as a self-organizing system (2.3).
In Section~3, we describe AutoCast, our approach for
adaptive data dissemination. Section~4 introduces the concept of
Hovering Data Clouds (HDCs), and describes how they can be used for
distributed recognition of traffic jams. Extensions for identifying
types and parameters are described in Section~5, along with
simulation results. Further extension of our ongoing work is
discussed in the concluding Section~6.

%    More and more people participate in traffic (cf.~\cite{statb06}),
    %the flow of traffic influences on air pollution and traffic jams
    %have impact on economic and social life. Thus, the interest in
    %possibilities to coordinate road users and to influence on them
    %to improve conditions like travel time, petrol consumption or
    %safety grows.
    %GGF. HIER NOCH WAS ZU ORGANISCHEN EIGENSCHAFTEN; WIE Z.B.
    %SELBSTORGANISATION.
    %This gives rise to a couple of questions.
   %ALLERDINGS AUCH SCHON TRAFFIC SIMULATION WEIT DAVOR (ZEITLICH).\\
%
%%%%%%%%%%%%%%%%%%%%%%%%%%%%%%%%%%%%%%%%%%%%%%%%%%%%%%%%%%%%%%%%%%%%%%%%%%%%%%%%%%%%%%%%%%
\section{Related Work}
%\section{Traffic, Self-Organization, and Communication}

\subsection{Communication Networks and Data Dissemination}

In today's communication world, wireless networks such as GSM in the
wide area and WLAN in the local area range have become ubiquitous.
Still, most applications using these networks rely on a thoroughly
managed infra\-structure such as, for instance, base stations in GSM
or access points in WLAN. Many research activities, however, already
go one step further and make use of the fact that more and more
mobile devices with radio communication capabilities are available.
These devices are not necessarily bound to communication
infrastructures, but can instead create a spontaneous network, a
so-called {ad-hoc network}, in order to allow communication among
some of the processors (and not necessarily to the outside). In its
sophisticated form, some of the processors in an ad-hoc network act
as relay stations and transport data from a source to a destination
that is not in direct radio range (multihop ad-hoc network). For an
overview on ad-hoc networks see the book by Perkins~\cite{p-ahn-01}.

%\subsection{Data Dissemination}
Unicast is the most popular way of communication and can be seen as
the standard for Client-Server based communication in the internet.
Nowadays, using wireless multi-hop networks many applications have
different requirements and various protocols and communication
approaches have been developed in the past to match new
applications' demands.

In sensor networks the communication paradigm shifts away from the
node-centric way where data is delivered between nodes identified by
addresses to a data-centric way of communication. The basic idea of
the data-centric communication is that nodes subscribe to a type of
data identified by a unique name and receive data associated with
this name as shown in \cite{638335,can,774785}.
%The ``name'' and the location of data may be stored in a hash map like done in peer-to-peer systems (see \cite{stoica01chord, rowstron01pastry}).
Since data is often sent to one or only few sinks in sensor
networks, approaches like \cite{YLC02} deal with moving sinks while
the rest of the network stays immobile. In any case, routes between
the originator of data and its subscribers are needed to transport
data through the multi-hop network. All these approaches fail for
the fast changing network topologies in vehicular ad-hoc networks
(VANETs).
%Furthermore, data subscribers need to know names of desired data.
%Another drawback is the insufficient handling of dynamically generated data in reaction to rapid changes.

%Another field where new approaches for data dissemination have been suggested recently are VANETs (see \cite{HBE01,cseh02mobile}).
Projects like \cite{FLEETNET} address the new challenges of VANETs
but often use data dissemination approaches limited to emergency
data. Like other examples as \cite{1023879,OKG2006} emergency
notifications are assumed to occur only rarely, statically and will
be short-lived. By contrast, our approach is able to handle numerous
data units in parallel, even when they are disseminated at the same
time to arbitrary directions and created at arbitrary positions in
the network.

Approaches that concentrate on disseminating traffic conditions,
like \cite{wischhof03adaptive,xuba2006}, focus on the adaptation of
broadcast interval, e.g., according to the vehicle's speed. The
proposed techniques are closely bound to specific applications with
fixed sized road segments and distinguish only between regular
communication and emergency data. In particular, they are not
designed for dynamic appearance and heterogeneity of data units with
individual life time and suddenly occurring long-lived data that
describes, e.g., a traffic jam.

A further optimization to save bandwidth while ensuring that every
node gets as much data as possible is described in \cite{313529}.
Each data unit is represented by a hash value. In a unicast
approach, new data is sent after a three way handshake comprising
advertisement, request, and delivery of data units.

Each approach has its individual optimum working condition and is
mainly created on the fly to solve a particular problem. In
Section~\ref{sec:protocol} we will derive step by step a generic
optimized protocol for data dissemination inspired by our AutoNomos
application.

\subsection{Traffic and Telematics}

As the interest in guiding and organizing traffic has been growing
over recent years, the scientific interest in traffic as a research
topic has developed quite dramatically. For an overview (``Traffic
and related self-driven many-particle systems''), see the excellent
survey~\cite{h-trsdmps-01}. Obviously, research on traffic as a
whole is an area far too wide for a brief description in this short
overview; we focus on a particular strain of research that is
particularly relevant for our proposed work, as it appears to be
most suited for simulation and extension to decentralized,
self-organizing systems of many vehicles.

It is remarkable that until the early '90s, efforts for simulating
traffic were based on complex multi-parameter models of individual
vehicles, resulting in complex systems of differential equations,
with the hope of extending those into simulations for traffic as a
whole. Obvious deficiencies of this kind of approach are manifold:

\begin{enumerate}
\item Because the behavior of even just an individual vehicle is guided
by all sorts of factors influencing a driver, the hope for a closed
and full description appears hopeless.

\item Determining the necessary data for setting up a simulation
for a relevant scenario is virtually impossible.

\item Running such a simulation quickly hits a wall; even with today's
computing power, simulating a traffic jam with a few thousand
individual vehicles based on such a model is far beyond reach.
\end{enumerate}

A breakthrough was reached when physicists started to use a
different kind of approach: instead of modeling vehicles with
ever-increasing numbers of hidden parameters, try to consider them
as systems of many particles, each governed by a very basic set of
rules. As Nagel and Schreckenberg managed to show \cite{ns-caft-92},
even a simple model based on cellular automata can produce
fractal-like structures of spontaneous traffic jams, i.e., complex,
self-organizing phenomena. Over the years, these models
\cite{n-hsmtf-95} were generalized to two-lane highway
traffic~\cite{rns-tltsc-96}, extended for simulating commuter
traffic in a large city \cite{rn-estsd-97}, and have grown
\cite{n-cata-02} to the point of being used for real-time traffic
forecasts for the 2250 km of public highways in the German federal
state of North Rhine-Westphalia \cite{kfc-osfnn-00,phc-tsann-03}
(see www.autobahn.nrw.de.) Also, see the book chapter
by Nagel \cite{n-tn-03}.

A closely related line of research uses an approach that is even
closer to particle physics; see \cite{nww-sfatf-03} for an excellent
overview of models for traffic flow and traffic jams, with about 150
relevant references. Among many others, particularly remarkable is
the approach by \cite{k-mmtfi-97}: this model reproduces properties
of phase transitions in traffic flow, focusing on the influence of
parameters describing typical acceleration and deceleration
capabilities of vehicles. This is based on the assumption that the
capabilities of drivers to communicate and coordinate are basically
restricted to avoid collisions, which until now is frustratingly
close to what drivers can do when stuck in dense traffic.

Parallel to the scientific developments described above, the
interest in and the methods for obtaining accurate traffic data has
continued to grow. The employment of induction loops and traffic
cameras has been around for quite a while, but despite of enormous
investments, e.g., 200 Mio.\ Euros by the German Federal Ministry
for Transport, Building and Urban Affairs (BMVBW) \cite{bmv-pvb-02} for
putting up systems for influencing traffic, the limits on tracking
individual vehicles, as well as following particular traffic
substructures are obvious. A more recent development is the use of
{\em floating car data:} By keeping track of the movements of a
suitable subset of vehicles (e.g., taxis in Berlin city traffic),
the hope is to get a more accurate overall image of traffic
situations, both in time and space \cite{kl-fcdau-00}. However, even
this approach relies on the use of the central processor paradigm,
and does not allow the use of ad-hoc networks for the active and
direct interaction and coordination between vehicles.

\subsection{Traffic as a Self-Organizing Organic System}

The structure of traffic is a phenomenon that is self-organizing at
several levels; see \cite{gh-wccsso-03} for a philosophical
discussion of self-organization in multi-level models. But even
though the behavior of and the interaction between motorists has
been observed for a long time, the possibilities arising from modern
technology allowing direct and decentralized complex interaction
between vehicles has hardly been studied. The only efforts we are
aware of combine game theory with traffic simulations. (For example,
see the symposium organized by traffic physicist Schreckenberg with
game-theory Nobel laureate Selten \cite{ss-hbtn-04}.) Neither make
use of mobile ad-hoc networks and distributed algorithms in large
networks.

Several research projects focus on enhancing traffic flow by distributed
methods. A model based on multi-agents systems is described in
\cite{1160653}. Depending on local congestion, a dynamic toll is charged to
influence the drivers' routing decisions. Camurri et al.~\cite{Camurri2006}
propose a distributed approach using Co-Fields for routing vehicles around
crowded areas on a urban grid-like road map. Techniques for detecting
traffic anomalies are also developed in centralized systems like the
'`System for Automatic Monitoring of Traffic'' as proposed in
\cite{Bandini2002}: Video cameras are installed along highways and traffic
events are derived and monitored over time based on the composed view of the
cameras.

All these approaches offer methods that try to solve certain problems arising
in the field of traffic management. Our approach offers a self-organizing
framework for decentralized applications; depending on the setting (urban,
highway) and the current traffic situation, different strategies can be
integrated to detect and overcome adverse traffic conditions.

\section{Data Dissemination}

\subsection{Protocol}
\label{sec:protocol}

%%%%%%%%%%%%%%%%%%%%%%%%%%%%%%%%%%%%%%%%%%%%%%%%%%%%%%%%%%%%%%%%%%%%%%%%%%%%%%%%%%%

%Vorüberlegungen
The proposed protocol \textit{AutoCast} aims at applications
that need to communicate in a many-to-many manner, without a need to
set up an association or connections between network nodes. In a
traffic information system, each car contributes to the knowledge of
road conditions that may be important for nearby cars.

In general, dissemination of data in mobile ad-hoc networks can be
achieved in two ways. This may either be by the movement of network
nodes (see \cite{grid:grossglauser01mobility}) or by multi-hop
ad-hoc communication between nodes. Because communication is much
faster than carrying data piggybacked on moving nodes, it is
preferred in most scenarios. However, node movement can support
communication when networks are partitioned, e.g., by using
opposite-lane traffic for bridging gaps between cars.

Because ad-hoc networks already use a broadcast medium, unicast
communication is an artificial constraint. Sending broadcast
messages is more efficient than unicast messages, gaining even more
with increasing network density. Even if a particular data unit is
not useful for a node, it can assist in further dissemination of the
data unit.

%%%%%% flooding

The most intuitive technique is pure \textit{flooding}, where each
node receiving a data unit rebroadcasts it exactly once and as soon
as possible. Flooding can be fast for fully connected networks.
However, the single-rebroadcast property causes network partitions
to stop data forever. As a consequence, flooding cannot cross partitions in the network; additionally, it jams the wireless channel in
dense networks with a broadcast storm \cite{NTC99}.

%%%%%% mile

A well-established method for disseminating data slowly and more
reliably, even when network partitions occur frequently, is a
periodic rebroadcast of received data with a short delay. As
described in \cite{1052873}, the protocol \textit{MILE} is designed
for the exchange of location information. We enhance \textit{MILE}
to work with generic data units instead of location information. By
randomly choosing several data units from all locally known data
units when broadcasting, data dissemination reaches an acceptable
speed. The main drawback is that this technique does not scale with
increasing network density and increasing number of data units in
the network.

It can easily be seen from the detailed results in
\secref{results} that both basic approaches have advantages and
disadvantages.

In order to measure the best possible performance, we introduce a
\textit{theoretical} protocol as benchmark. This protocol assumes
unlimited transmission rate, propagation speed of light, and a
perfect intuition of the sender as to which data units need to be
sent to which nodes, just in the moment when they are able to
receive them correctly. This happens magically, especially when
network partitions merge again without any delay and additional
communication overhead.

As a first improvement, we optimize \textit{MILE} by reducing the
amount of data that needs to be transmitted periodically. The idea
is to use simple and well-known hash values. Nodes create short hash
values from data units, so-called IDs---sometimes also called
metadata---and send these instead of complete data units. This
complete list of IDs is broadcast periodically by each node,
 together with a subset of data units.
If a node gets an incomplete list of IDs from a neighbor, it will
add the missing data units in its next update packet. By this simple
extension we avoid an explicit request of missing data by individual
nodes and thereby achieve an additional reduction of bandwidth
usage. The drawback is an increased delay, as nodes add the content
of the data units only when other nodes within the transmission
radius are found that do not know about a particular data unit. We
call this extension \textit{MILE on-demand}.

We combine the ideas of \textit{flooding} and \textit{MILE
on-demand} for further reducing the communication overhead and
increasing the speed of data dissemination, as well as the data
delivery ratio. We call the new protocol \textit{AutoCast}; it works
as follows, making use of two basic mechanisms.

Newly generated data units are flooded through the network in the
beginning, but only a portion of the nodes participate actively in
the flooding. Instead of using the magic numbers of 60\% to 80\% as
a forwarding probability (as suggested in \cite{867194}), we adapt
to the dynamics and irregularity of the network. Nodes derive the
probability from their number of neighbors. To avoid broadcast
storms, on average only two nodes of those receiving a new data unit
rebroadcast it. It has been shown in \cite{hellFisICWN02} that on
the average only about 40~\% of the neighboring nodes receive the
data unit for the first time as 60~\% of the nodes have received the
previous broadcast already. Consequently, a node with 10 neighbors
forwards a data unit with a probability of $2/(10\cdot0.4)=0.5$,
which is according to the results of \cite{867194} for this
scenario.
However, the forwarding probability for single nodes will decrease further when network density increases, thus ensuring scalability. In a traffic jam, the number of neighbors can reach 100 cars easily where with our approach an individual node forwards data unit with a probability of $0.05$.% with \textit{AutoCast} which is more than 10 times better than suggested by \cite{867194}.

The second mechanism was introduced by \textit{MILE on-demand}:
Periodically rebroadcast IDs of elder data units, because due to bad
luck, flooding might stop sometimes when several nodes do not
forward data. Periodic rebroadcasts are also important to reach
locally consistent states in the network, especially when new nodes
join the network or network partitions merge. Like the forwarding
probability, the rebroadcast interval also depends on the number of
neighbors, and in addition on the network dynamics. In a static
network in which nodes neither move nor appear, periodic
rebroadcasting does not help at all, as flooding already delivered
the data units to all reachable nodes. On the other hand, increasing
speed of nodes combined with frequent network partitioning will
force the update interval to reach zero, which means all nodes
broadcast permanently. Thus, the key issue is to determine the
optimal update interval; furthermore, how can individual nodes
calculate it on their own?

Assume that we know $n$ as the size of a node's one-hop
neighborhood. The waiting time until the next rebroadcast is
calculated as $n/p_{\mathrm{ref}}$, where $p_{\mathrm{ref}}$ is a
constant that describes the desired number of broadcasts per second.
We will explain our choice of $p_{\mathrm{ref}}$ more detailed in
\secref{results_autocast}. A positive side-effect is the following:
Cars driving near a network partition boundary send twice the number
of packets as cars driving inside that partition, as border nodes
have only half the expected number of neighbors.

%%%%%%%%%%%%%%%%%%%%%%%%%%%%%%%%%%%%%%%%%%%%%%%%%%%%%%%%%%%%%%%%%%%%%%%%%%%%%%%%%%%
%%%%%%%%%%%%%%%%%%%%%%%%%%%%%%%%%%%%%%%%%%%%%%%%%%%%%%%%%%%%%%%%%%%%%%%%%%%%%%%%%%%

\subsection{Simulation Setup}
\label{sec:setup}

%%%%%%%%%%%%%%%%%%%%%%%%%%%%%%%%%%%%%%%%%%%%%%%%%%%%%%%%%%%%%%%%%%%%%%%%%%%%%%%%%%%

After having discussed five different approaches (including the
\textit{AutoCast} protocol), we set up a simulation environment to
evaluate and compare them.

We have chosen a dynamic highway scenario with varying
network density and the influence of opposite-lane traffic for our
protocol's performance. Cars drive on a highway section of 10 km, 
with two lanes in each direction and an average speed of
100~km/h. In order to reach realistic node movements that will
appear in VANETs due to individual cars' behavior, we used
the traffic simulator SUMO (see \cite{Krajzewicz_et_al2002_1}),
which is based on the microscopic car following model described in
\cite{k-mmtfi-97}. The mean distance between two consecutive cars on
one lane is around 110~meters, leading to an overall
mean car density of 36~cars/km. Because road density is hard to
compare to other simulation setups, \tabref{car_density} shows
neighborhood sizes in our setup that result from different fractions
of cars equipped with AutoNomos devices, so-called {\em penetration
rates}.

\begin{table}[tp]
\setlength{\tabcolsep}{0.81mm}
\caption{\small Average neighborhood sizes for different penetration
rates.} \label{tab:car_density} \centering
% Some packages, such as MDW tools, offer better commands for making tables
% than the plain LaTeX2e tabular which is used here.
\begin{tabular}{|l|c|c|c|c|c|c|c|c|c|c|c|}
\hline
penetration rate [\%] & 5 & 10 & 20 & 30 & 40 & 50 & 60 & 70 & 80 & 90 & 100 \\
\hline
neighborhood size & 0.9 & 1.8 & 3.6 & 5.4 & 7.2 & 9 & 10.8 & 12.6 & 14.4 & 16.2 & 18 \\
\hline
\end{tabular}
\end{table}

The nominal duration of our traffic simulation is 26~min, with an
initial startup time of 10~min to spread the cars all over the road.
The last 16~min of the generated cars' mobility are stored into
ns2-trace files, each with a different penetration rate.

ns-2 \cite{ns2} is used as network simulator for performance
evaluation of the different data dissemination protocols. All
simulations use standard IEEE 802.11~MAC-layer, with a radio range
of 250~m and a bandwidth of 1~Mbps in combination with the Two-Ray
Ground propagation model. Periodically, the car driving closest to
km~5 at the appointed time generates a data unit, which is disseminated
over the simulated road (5~km in each direction); the unit's lifetime
is set to 50 seconds.

Each protocol is simulated with different penetration rates, as
shown in \tabref{car_density}, and between two and 50 data units
that need to be disseminated concurrently.

%%%%%%%%%%%%%%%%%%%%%%%%%%%%%%%%%%%%%%%%%%%%%%%%%%%%%%%%%%%%%%%%%%%%%%%%%%%%%%%%%%%
%%%%%%%%%%%%%%%%%%%%%%%%%%%%%%%%%%%%%%%%%%%%%%%%%%%%%%%%%%%%%%%%%%%%%%%%%%%%%%%%%%%

\subsection{Results}
\label{sec:results}

%%%%%%%%%%%%%%%%%%%%%%%%%%%%%%%%%%%%%%%%%%%%%%%%%%%%%%%%%%%%%%%%%%%%%
\begin{figure*}[!t]
        \begin{center}
          \mbox
          {
            \subfigure[Radio channel usage per km, 2 simultaneous data units]
            { \includegraphics[width=0.45\columnwidth,height=4cm]{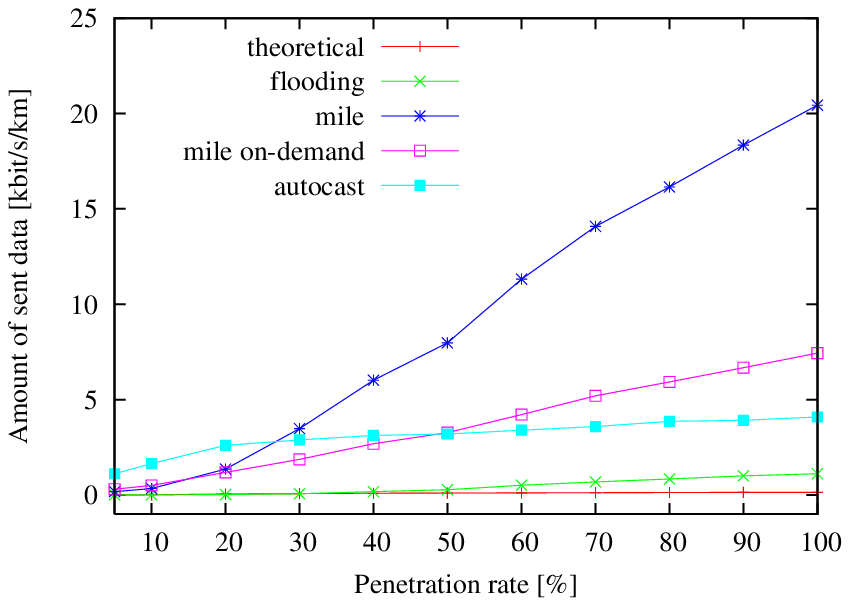} \label{fig:datarate_2_data_units} }
            \quad
            \subfigure[Radio channel usage per km, 50 simultaneous data units]
            { \includegraphics[width=0.45\columnwidth,height=4cm]{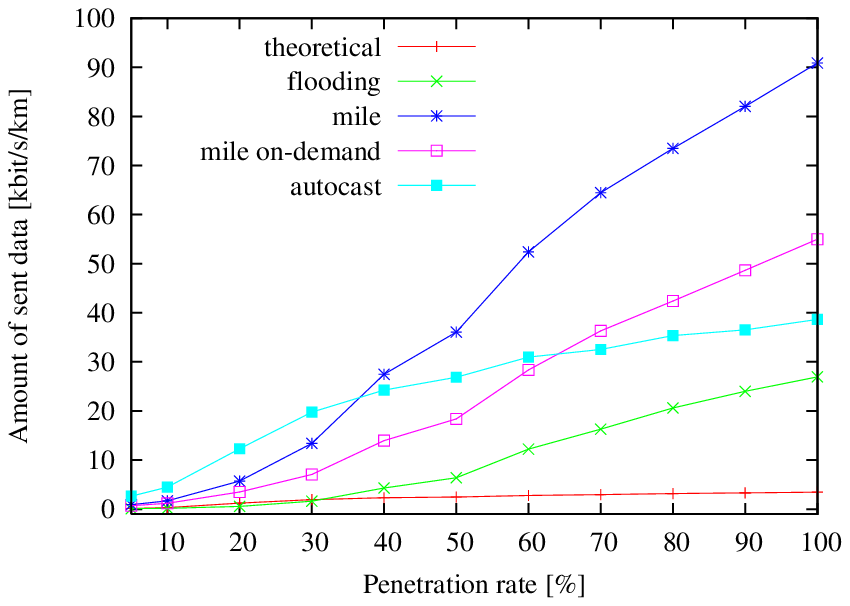} \label{fig:datarate_50_data_units} }
          }

          \vspace*{0.2cm}

          \mbox
          {
            \subfigure[Speed of data traffic, 2 simultaneous data units]
            { \includegraphics[width=0.45\columnwidth,height=4cm]{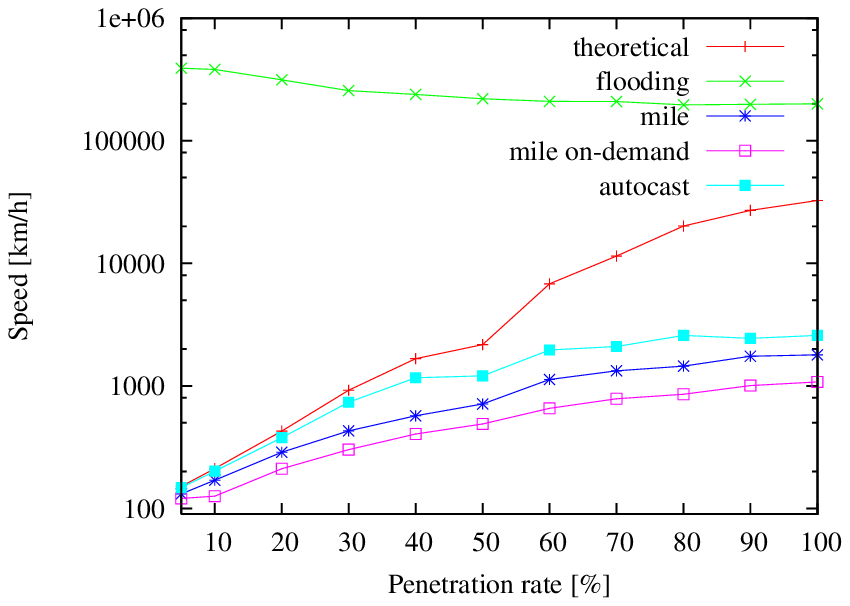} \label{fig:dataspeed_2_data_units} }
            \quad
            \subfigure[Speed of data traffic, 50 simultaneous data units]
            { \includegraphics[width=0.45\columnwidth,height=4cm]{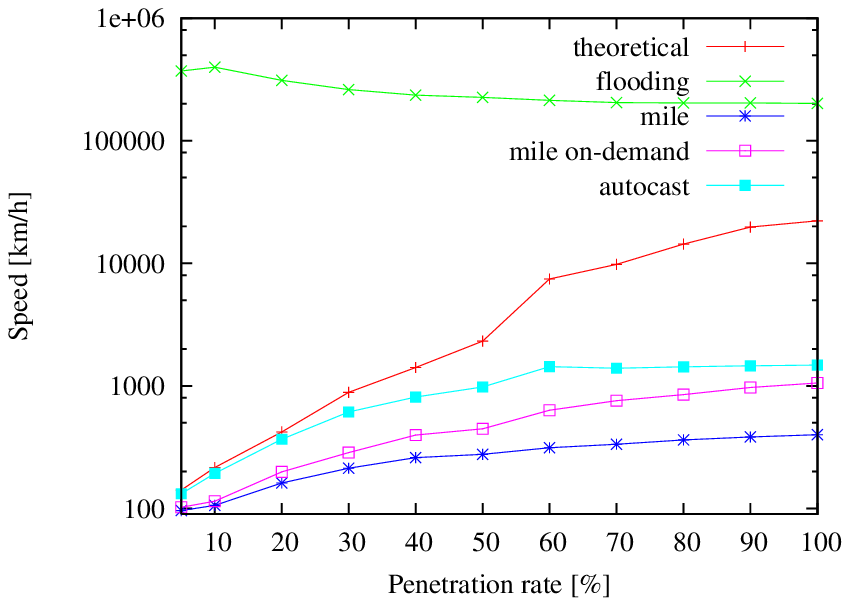} \label{fig:dataspeed_50_data_units} }
          }

          \vspace*{0.2cm}

          \mbox
          {
            \subfigure[Data delivery ratio, 2 simultaneous data units]
            { \includegraphics[width=0.45\columnwidth,height=4cm]{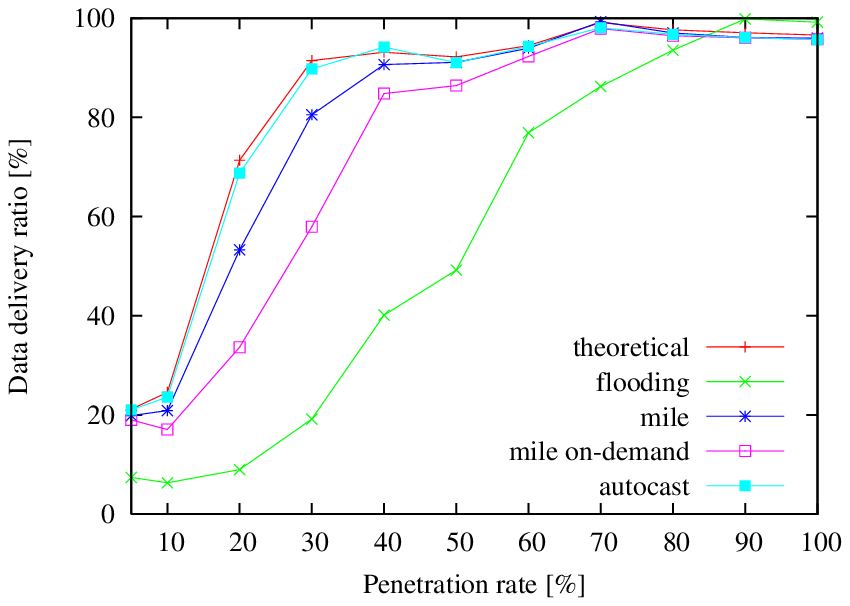} \label{fig:informed_2_data_units} }
            \quad
            \subfigure[Data delivery ratio, 50 simultaneous data units]
            { \includegraphics[width=0.45\columnwidth,height=4cm]{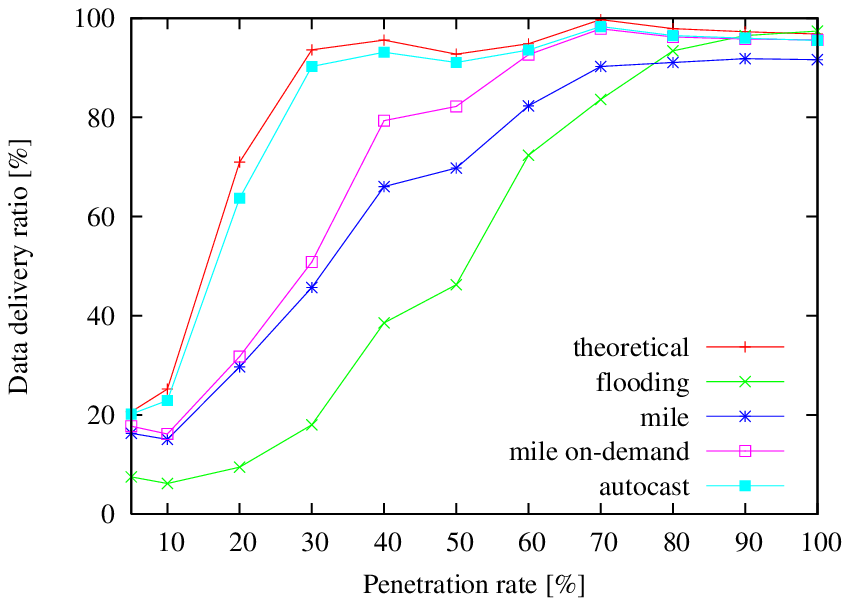} \label{fig:informed_50_data_units} }
            }

\vspace*{-0.1cm}

          \caption{\small Simulation results comparing the different data dissemination        \label{fig:algorithm_compare}
algorithms for few data units (left column) and 50 data units (right
column).}
        \end{center}
\vspace*{-0.6cm}
\end{figure*}

%%%%%%%%%%%%%%%%%%%%%%%%%%%%%%%%%%%%%%%%%%%%%%%%%%%%%%%%%%%%%%%%%%%%%%%%%%%%%%%%%%%

\figref{algorithm_compare} shows the results of the simulation, with
each line in the graph showing a protocol. In all plots the $x$-axis
shows the penetration rate of cars that participate in the VANET.
The left column shows the protocols' behavior, if only two data
units are disseminated. Figures on the right show the results for 50
concurrent data units.
In order to leave enough network capacity for other applications and
protocols, data dissemination should be optimized for low bandwidth;
Figures \ref{fig:datarate_2_data_units} and
\ref{fig:datarate_50_data_units} show the transmitted data per km,
as concurrent communication is possible if sending nodes have a
distance of more than four times the transmission radius.
Figures \ref{fig:dataspeed_2_data_units} and
\ref{fig:dataspeed_50_data_units} show the achieved speed of data
dissemination. A fast speed is preferable, because data units may
comprise time-critical data like emergency messages.
To evaluate the success of data dissemination, Figures
\ref{fig:informed_2_data_units} and \ref{fig:informed_50_data_units}
present the amount of successfully delivered data units.

%\subsection{Theoretical protocol}

The \textit{theoretical} protocol sends a broadcast only if it will
successfully inform a car, so less than 4~kbit/s/km of bandwidth are
consumed in any case. With low penetration rate, the bounding factor
for data speed is almost completely the cars' driving speed. As
expected, it rises with increasing penetration rate, up to more than
20000~km/h. This speed cannot be achieved by any other protocol, as
in reality there is a trade-off between data speed, data delivery
ratio and rebroadcasting interval. The data delivery ratio shows
that a reasonable usefulness can be achieved with a minimal
penetration rate of 30~\% standing for 10.8 equipped cars per km.
With a further increase in the number of cars, the ratio of
delivered data units grows only marginally.

%\subsection{Flooding protocol}

At first sight the \textit{flooding} protocol performs surprisingly
well. It consumes few bandwidth and achieves a speed of above
100,000~km/h. However, the poor data delivery ratio puts that result
in the right perspective, as with pure flooding a data unit will stay
in its network partition. Consequently, flooded data is delivered
either very fast or never.

%\subsection{MILE protocol}

With regard to enhancing the data delivery ratio, in particular in
the case of low penetration rates, the \textit{MILE} protocol
achieves a remarkable improvement, approaching the theoretical
results. If more data units need to be disseminated than what fits into
one broadcast packet, the achieved data speed decreases from nearly
1800~km/h to under 400~km/h, even in case of a 100~\% penetration
rate. Moreover, the data delivery ratio drops as well.

%\subsection{MILE on-demand protocol}

Due to the exchange of data unit IDs, the protocol \textit{MILE
on-demand} can suppress the rebroadcasting of full data units that
are already known by cars in the direct vicinity. The drawback of
this method is a slight decrease in data speed, because the sender
needs to know about missing data units before delivering them. Due
to this effect \textit{MILE on-demand} performs worse than pure
\textit{MILE} in case of only few data units. Nevertheless, the
protocol's performance remains stable if more data units need to be
handled. So far all protocols use fixed broadcast intervals of 2s.
This results in a linear increase of bandwidth usage when
 more cars participate in the VANET.

\label{sec:results_autocast}

As mentioned in \secref{protocol}, \textit{AutoCast} produces a
constant number of broadcast packets per second
($p_{\mathrm{ref}}$), no matter how many nodes generate them. In
order to calculate $p_{\mathrm{ref}}$ for our scenario, we analyze
the \textit{MILE on-demand} curve and find a minimum of 60~\%
penetration rate for a data delivery ratio above 90\%. With 10.8
neighbors, i.e., $p_{\mathrm{ref}}=10.8$ cars in the neighborhood
per 2~s, about 5~packets/s are transmitted. The value of
$p_{\mathrm{ref}}$ is a good choice for our scenario, but is
definitely not the optimum for all ad-hoc networks. This parameter
needs to be analyzed in more detail; we will address this problem in
future work. Nevertheless, bandwidth consumption remains stable,
independent of the network density, and depending only on the number
of concurrent data units. The data speed reaches about 2000~km/h,
enough to cross Germany in less than 30~min. The data delivery ratio
gets close to the \textit{theoretical} protocol, so even the primary
goal of reaching as many cars as possible is achieved.

\AutoCast clearly outperforms the other protocols and gets close to
the theoretical maximum with respect to data dissemination speed and
data delivery ratio. Due to a limited network overhead, it leaves
enough room for additional applications and protocols in the ad-hoc
network.

\section{Hovering Data Clouds for Traffic Jams}
\subsection{Hovering Data Clouds}

A dynamically changing system such as a traffic jam consists of many
ever-changing objects, as cars located at different positions keep
moving with respect to back and front of the queue. If we want to
maintain useful information related to the back of a traffic jam, we
have to keep shifting the related roles from one car to the next,
together with all relevant data. Thus, we look for a {\em local}
data structure with the following properties:

\begin{itemize}
\item The data structure self-organizes with the onset of a traffic
jam, and it ceases to exist when the jam disappears.
\item It is located at a useful virtual location, which
is defined by the traffic jam, e.g., its back.
\item The structure continues to exist, even as their
current carriers move or change their role.
\item It contains up-to-date information that describes the traffic jam.
\end{itemize}

We call such a structure a {\em Hovering Data Cloud} (HDC). In this
section, we describe how to deal with the above properties in the
context of a relatively simple traffic scenario. Even though this presentation
focuses on the context of traffic jams, there are a number of other
scenarios that give rise to HDCs:

\begin{itemize}
\item Keeping track of large swarms of moving animals poses similar
challenges of dynamically updating information, without relying on a
fixed host. %(See \cite{Butler04bbis} for a scenario in the
%context of cows.)
\item Pedestrian traffic also consists of a large number of
individuals, each capable of (and nowadays routinely)
carrying a mobile device. The resulting patterns can be even more
complicated than  highway traffic; see \cite{helbing} for
an introduction. Quite clearly, HDCs can be used for a number
of different pedestrian scenarios and applications.
\item Even for scenarios with devices at fixed locations,
HDCs may be useful: When a mobile object is tracked by a sensor
field, historic information is accumulated over time that is not attached
to the object (as it may not be equipped with an electronic device),
but also not present at each individual sensor.
Passing this information from host to host along the tracking path
amounts to maintaining an HDC.
\end{itemize}

\begin{figure*}[!t]
        \begin{center}
                \includegraphics[width=\columnwidth] {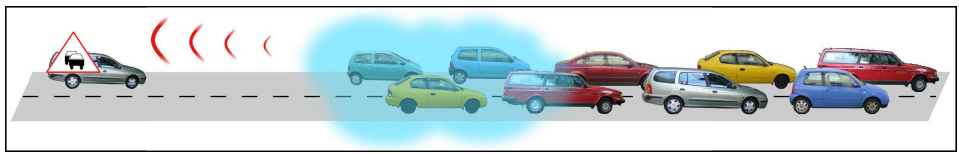}
                \caption{\small A Hovering Data Cloud at the back of a traffic jam informs incoming vehicles.}
                \label{fig:traffic_jam_end}
        \end{center}
\end{figure*}

\figref{traffic_jam_end} depicts a traffic scenario in which an HDC is
already located at the end of a traffic jam. Vehicles approaching
the HDC, participate in storing, and enrich the data by monitoring
the underlying phenomenon. Furthermore, the data of the HDC is
propagated and sent to oncoming vehicles as warning messages.
Obviously, there are two reasons for communication between nodes in
this example: firstly, vehicles holding the HDC communicate in order
to maintain the HDC (intra-HDC communication); secondly, vehicles
outside the HDC's area need to get informed about the HDC (inter-HDC
communication).

We started our work on this concept by considering stationary HDCs
\cite{wshff-hdcdsois-06}; these arise when performing measurements
at predefined locations. Using the simulator SUMO (Simulation of
Urban MObility) \cite{Krajzewicz_et_al2002_1} for generating
mobility traces, and ns2 \cite{ns2} for network simulation, we were
able to give a good reproduction of actual traffic 5~km away from
the HDC. However, the main interest and justification for HDCs
arises in more dynamic situations, in which HDCs may not remain
stationary. This work is described in detail in the rest of this
section.

\subsection{Scenario}
\begin{figure}[tp]
\centering
\epsfig{file=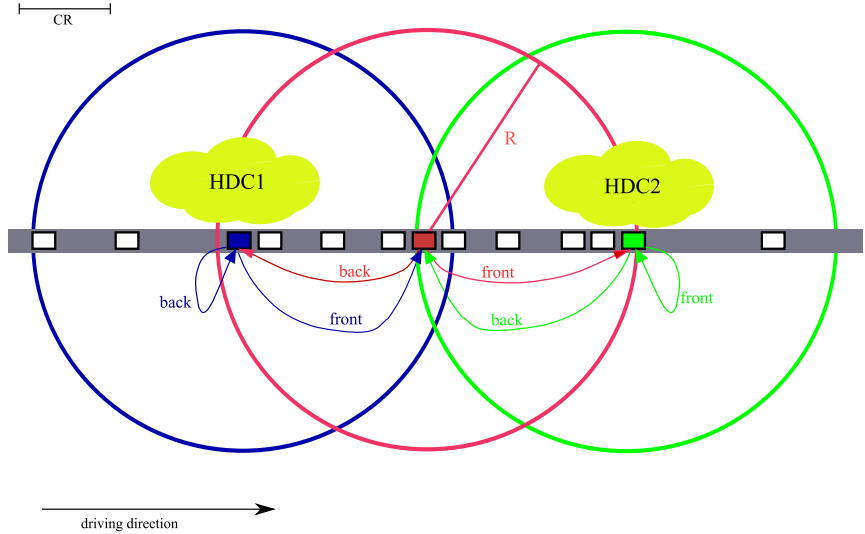, width=.9\textwidth}\\
\caption{\label{cars}\small Example for the two HDCs: different cars
(rectangles) determine front and back (assumption: all drive sufficiently
slowly), HDCs arise if all further conditions are met. }
\end{figure}

We consider a single-lane highway. For convenience, we assume that
cars move from left to right, as shown in Figure~\ref{cars}, i.e.,
positions with lower coordinates are shown on the left. Any vehicle
carries a computing and wireless communication device, each with a
unique identifier, and has a reliable way of measuring time and
location, e.g., by using GPS. Cars can communicate if they are
within broadcast range of each other; this range is denoted by $R$.
Communication delays are relatively small, we assume 10\,ms as
transmission delay.

Now we examine traffic patterns. Depending on speed and traffic
density, a traffic jam may form. This gives rise to two HDCs, one
at the front and and one at the back of the queue, cf.\ Figure~\ref{cars}; these HDCs are maintained while the jam continues to
exist. In the following we describe the details of how this can be
achieved; note that the same basic variables and processes are
maintained in each processor.

\subsection{Variables}

We will make use of the following variables and parameters for each
processor.

%%%%%%%%%%%%%%%%%%%%%%%%%%%%%%%%%%%%%%%%%%%%%%%%%%%%%%%%%%%%%%%%%
% Aenderung in diesem Abschnitt (Idle nodes, die active werden)
%%%%%%%%%%%%%%%%%%%%%%%%%%%%%%%%%%%%%%%%%%%%%%%%%%%%%%%%%%%%%%%%%

{\em status}$_i$ (with possible values idle, joining, active)
describes the current state of a processor, where the index $i \in
\{$back, front$\}$ refers to the HDCs marking back and front of the
traffic jam; initially, all processors are idle. In the absence of a
traffic jam, idle processors near an HDC become active
immediately. If a jam exists, processors become joining if they are
within reach of the current HDC, described by a radius
$r_{\mathrm{HDC}}$ around an HDC's current location.
An active processor becomes idle, if it ceases to be near the
boundary of the
jam, i.e., if its distance to an HDC position exceeds
$r_{\mathrm{HDC}}$. Note that the status joining is only necessary
in the presence of information that is not known to all processors.

{\em state} describes the values of all variables stored on a
processor.

The coordinates {\em location}$_\mathrm{back}$ and {\em location}$_\mathrm{front}$
describe the current positions of an HDCs that mark back and front of
the traffic jam. For clarity we should mention that the variables
\textit{loc, v} and ID refer to the actual processor, \textit{l, g}
and \textit{ident} refer to the processor from which the actual
message \textit{m} was received. {\em buffer} is used for storing
arriving messages. {\em clock} is used for keeping track of real
time. {\em env} is a table for storing the data of all broadcasting
processors within distance $R$ (maximum size: $3 \times \lceil
2R/(\mbox{minimum distance})\rceil$).

If a processor has a left neighbor (i.e., a following car) within
congestion radius {\em CR}, the flag $p$ is set to 1. Analogously, $q=1$
indicates a right neighbor (i.e., a preceding car) within congestion
radius {\em CR}.

The auxiliary variables {\em ahead, congestion\_counter,
congestion\_participant, congestion\_participant\_before, s,
q\_before, p\_before, P$\_v$, Q$\_v$, p$\_v$, q$\_v$, t, front,
back\_id, front\_id} are all initialized with 0; {\em back} is
initialized with the largest possible value of a location on the
road.

If status joining is necessary, then $v_{0_\mathrm{back}}$,
$v_{0_\mathrm{front}}$ are the initial states of an HDCs.

\begin{figure}[tp]
\centering
\epsfig{file=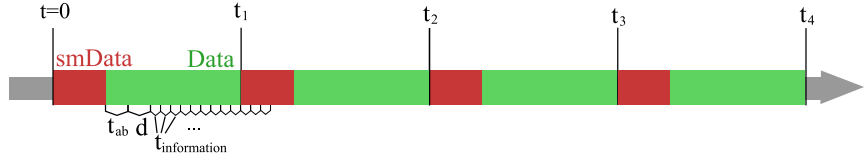, width=.9\textwidth}\\
\caption{\label{time}\small Example for time slots and timer. }
\end{figure}

$t_{\mathrm{data}}$, $t_{\mathrm{smdata}}$,
$t_{\mathrm{information}}$, $t_{\mathrm{aheadInfo}}$ are times that
are used for updating the indexed variables. $t_{\mathrm{data}}$,
$t_{\mathrm{smdata}}$ occur alternately. $t_{\mathrm{information}}$,
$t_{\mathrm{aheadInfo}}$ describe times with evenly spaced
intervals. The function settimer is used with two arguments: the
first one indicating a certain point in time, the second one the
callback function {\it f} which is evoked at this point in time
(with {\it onTimer(f)}). A possibility to describe a point in time
is to use the function next-multiple: Given a variable that
indicates the time divided into intervals of equal length,
next-multiple gives the next point in time where such an interval
starts. Consequently, using the function settimer with the argument
next-multiple(\textit{t}$_x$) ($x \in \{$information, aheadInfo$\}$)
sets the timer on the next time of \textit{t}$_x$ (even if some time
elapsed since the last timer was set), cf. Figure~\ref{time}.
% these updates occur at
%evenly spaced time intervals.
$t_{\mathrm{ab}}$ is the time period that passes until all
processors within $2R$ have started their {\em Data} interval.

A parameter that is related to the presence of a traffic jam is the
{\em congestion radius} {\em CR} that is a function of the current speed
of vehicles; if two cars violate this critical distance, it may be
an indication of a traffic jam.

Conversely, the {\em congestion velocity CV}, considers the velocity
in relation to the current distance; if two cars move more slowly than
appropriate for their current distance, this may also indicate a
traffic jam. (Note that this second parameter filters out cars that
tailgate at high speed. As pointed out by \cite{brilon05}, it works
best to consider speed for recognizing traffic jams.)

Finally, $d$ is the critical bound on the latency in {\em LBcast}.

%%%%%%%%%%%%%%%%%%%%%%%%%%%%%%%%%%%%%%%%%%%%%%%%%%%%%%%%%%%%%%%%%%%
% Abbildung in Farbe - inkspace
%%%%%%%%%%%%%%%%%%%%%%%%%%%%%%%%%%%%%%%%%%%%%%%%%%%%%%%%%%%%%%%%%%%

%%%%%%%%%%%%%%%%%%%%%%%%%%%%%%%%%%%%%%%%%%%%%%%%%%%%%%%%%%%%%%%%%%%
% Abbildung in Farbe - ipe und andere Fahrtrichtung
%%%%%%%%%%%%%%%%%%%%%%%%%%%%%%%%%%%%%%%%%%%%%%%%%%%%%%%%%%%%%%%%%%%

%\begin{figure}[tp]
%\centering
%\epsfig{file=autos.eps, width=.9\textwidth}\\
%\caption{\label{cars}\small Example for the two HDCs: different cars
%(rectangles) determine front and back (assumption: all drive slow
%enough), HDCs arise if all further conditions are met. }
%\end{figure}

%%%%%%%%%%%%%%%%%%%%%%%%%%%%%%%%%%%%%%%%%%%%%%%%%%%%%%%%%%%%%%%%%%%
% Abbildung in Schwarz-weiß
%%%%%%%%%%%%%%%%%%%%%%%%%%%%%%%%%%%%%%%%%%%%%%%%%%%%%%%%%%%%%%%%%%%

%\begin{figure}[h]
%\centering
%\epsfig{file=autos-sw.eps, width=7.12cm}\\
%\caption{\label{cars}\small Example for the two HDCs: different cars
%(rectangles) determine front and back (assumption: all drive slow
%enough), HDCs arise if all further conditions are met. }
%\end{figure}

%%%%%%%%%%%%%%%%%%%%%%%%%%%%%%%%%%%%%%%%%%%%%%%%%%%%%%%%%%%%%%%%%%%
% Abbildung in Farbe - inkspace
%%%%%%%%%%%%%%%%%%%%%%%%%%%%%%%%%%%%%%%%%%%%%%%%%%%%%%%%%%%%%%%%%%%

%%%%%%%%%%%%%%%%%%%%%%%%%%%%%%%%%%%%%%%%%%%%%%%%%%%%%%%%%%%%%%%%%%%
% Abbildung in Farbe - ipe
%%%%%%%%%%%%%%%%%%%%%%%%%%%%%%%%%%%%%%%%%%%%%%%%%%%%%%%%%%%%%%%%%%%

%\begin{figure}[tp]
%\centering
%\epsfig{file=Zeitpunkte-bunt.eps, width=.9\textwidth}\\
%\caption{\label{time}\small Example for time slots and timer. }
%\end{figure}

%%%%%%%%%%%%%%%%%%%%%%%%%%%%%%%%%%%%%%%%%%%%%%%%%%%%%%%%%%%%%%%%%%%
% Abbildung in Schwarz-weiß
%%%%%%%%%%%%%%%%%%%%%%%%%%%%%%%%%%%%%%%%%%%%%%%%%%%%%%%%%%%%%%%%%%%

%\begin{figure}[h]
%\centering
%\epsfig{file=Zeitpunkte-sw.eps, width=7cm}\\
%\caption{\label{time}\small Example for time slots and timer. }
%\end{figure}

\subsection{Algorithm Description}
We give details of the algorithmic steps; see Algorithm 1 (Appendix
A).
%\ref{alg}.
%; seeFigures~1--8 for
%details.

We consider a discrete sequence of time slots, $t_i, t_{i+1},
\ldots$. The interval between two consecutive time slots is divided
into two subintervals: a small interval ({\em {smData}}), and a bigger
one ({\em Data}), cf. Figure \ref{time}. Thus, at the end of {\em
smData (onTimer(smData))} the timer for {\em Data} is initiated and
vice versa.

%%%%%%%%%%%%%%%%%%%%%%%%%%%%%%%%%%%%%%%%%%%%%%%%%%%%%%%%%%%%%%%%%
% Aenderung in diesem Abschnitt (Bei LBrecv(m))
%%%%%%%%%%%%%%%%%%%%%%%%%%%%%%%%%%%%%%%%%%%%%%%%%%%%%%%%%%%%%%%%%

In {\em onTimer(smData)} the current processor position and its
velocity are only broadcast within {\em CR}. If a processor receives such a
message, it is treated in subroutine {\em onTimer(NewMessage)} (as
all received
messages)---after an additional delay of $d$ (see {\em LBrecv($m$)}%,
%Figure \ref{fig:rec}
). In case a processor receives such a message from ahead, we set
$p$ equal to 1. Analogously, a message from behind results in
setting $q$ equal to 1.

For sending more data ({\em position, velocity}, $p$, $q$) in a
wider range in {\em onTimer(Data)} we distinguish several
situations. Only if the processor is participating in an HDC or if it
is neither caught up in a traffic jam ($t_i$) nor was so before
($t_{i-1}$) but falling below a certain velocity, it will send such
a message. This enables us to reduce the amount of transmitted
messages. However, non-active processors inside of a traffic jam or
with sufficient high speed do not send, the information is either
not important yet or the processor is not a potential participant of
a traffic jam.

Describing the consequences of being a congestion participant, we
need to consider how a processor achieves this situation. A
processor receives messages of type \textit{Data} from its
surrounding processors within $R$. The data of each such processor
is stored in \textit{env}. Afterwards, it is checked whether the
sending processor is located close (less than {\em CR}) to the receiver
and if the velocity is sufficiently low, e.g., clearly below 60 km/h
(see \cite{brilon05,k-mmtfi-97}.) If so, the back of the congestion
is computed from
the position of the two processors and the previous back position.
Furthermore, the processor becomes a participant of the traffic jam.
Similarly we check for all processors in \textit{env} whether they
are located close to the sending processor and fall below a velocity
of {\em CV}. In both cases we increment a counter for the congestion
(indicating a positive number of vehicles).

If the counter was incremented, the processor is close (less than
$r_\mathrm{HDC}$) to the back of the congestion, the back position
has no left neighbor (thus,
it is really the back) and all messages from processors within the
range of $R$ were received, then {\em Congestion} is invoked. The
front of the jam is treated analogously; {\em CongestionAhead} is
invoked here. If an HDC at the back has yet to receive information
from an HDC at the front, then the position of {\em front} is set
to the position of the most advanced processor within $R$.

The status of a processor is maintained in {\em Congestion}: if it
is active, we set a timer for {\em Information}, i.e., as long as
the processor is active, the position of an HDC at the back of the
jam, the position of an HDC at the front, and the current speed of
the back are broadcast. Only if the position of the back HDC has a
value greater than the current processor location, the processor
continues to broadcast, and processors approaching this position
become {\em joining}.

In {\em CongestionAhead}, {\em status}$_\mathrm{front}$ is updated;
if the processor is active, the timer for {\em AheadInfo} is set.
This means that the position of the front HDC is broadcast
regularly, as long as the processor is active. When such a message
{\em hdcdistance} is processed, the back HDC variable {\em front}
indicating the position of the front HDC is updated for an active
processor: inactive processors between front and back HDC pass on
the message towards the back.

%%%%%%%%%%%%%%%%%%%%%%%%%%%%%%%%%%%%%%%%%%%%%%%%%%%%%%%%%%%%%%%%%
% Aenderung in diesem Abschnitt
%%%%%%%%%%%%%%%%%%%%%%%%%%%%%%%%%%%%%%%%%%%%%%%%%%%%%%%%%%%%%%%%%

With increasing distance, clustering and updating data is performed
by the HDC transportation layer.

Thus, messages are broadcast to:
\begin{itemize}
\item relate the positions of the processors (messages broadcast in
{\em onTimer(smData), onTimer(Data)}, processed in {\em onTimer(NewMessage)}),
\item transmit the information of the HDC at the front of the
traffic jam to the one at the back of the congestion (messages in
{\em onTimer(AheadInfo)}),
\item transmit the information of the back HDC to the following cars
(messages in {\em onTimer(Information)}).
\end{itemize}

\subsection{Further Aspects of HDCs}
There are various extensions and generalizations of the above
scenario. An obvious next step is to extend our ideas to two-lane
highways (possibly stretching over several exits and even highway
crossings), or more refined HDCs that reflect substructures in a
traffic jam; other extensions and variants include the recognition
of bottlenecks in traffic, e.g., caused by a convoy of slow trucks,
accidents or emergency vehicles.

A qualitatively more challenging step is required when considering
more advanced structures that consist of several HDCs: an HDC simply
marks front or back of a traffic jam, but eventually we are
interested in more complex interaction between all involved
vehicles, e.g., when trying to smooth out complex stop-and-go
patterns, mark advisable exits for following cars, or even map
possible detours. We call such high-level structures {\em Organic
Information Complexes} (OICs). The underlying idea is presented in
\secref{autonomos_oic}. Details will be pursued and discussed in
future work.

\subsection{Simulation}

We have implemented our method, using the traffic simulator SUMO
(Simulation of Urban MObility, \cite{Krajzewicz_et_al2002_1}) for
generating traces of vehicular movement, and our own large-scale
distributed network simulator Shawn \cite{kpbff-snaswsn-05} for
handling the algorithmic side. Enhanced by an OpenSG-based 3D
visualization plug-in for Shawn by Baumgartner
\cite{b-reafp-06}, the resulting simulation can produce
videos; see Figure~\ref{isola} for some screen
shots. The initial state was chosen such that traffic jams emerge.
Starting with these trace files, SUMO controls the movement of the
vehicles (according to its internal rules, e.g., using the
car-following model of \cite{k-mmtfi-97}). The location and velocity
of all cars are written in an input file for Shawn. Thus,
Shawn only updates the location and speed and its focus is on the
management of messages. In Figure~\ref{isola} the resulting HDC at
the front of the traffic jam is marked in brown (with corresponding
cars labeled in yellow), while the HDC at the end is marked in red
(as are the corresponding cars.) This shows that
our methods do indeed produce the desired results.

%GROESSE???
%%%%%%%%%%%%%%%%%%%%%%%%%%%%%%%%%%%%%%%%%%%%%%%%%%%%%%%%%%%%%%%%%%%%%%%
\begin{figure}[tp]
\centering \epsfig{file=screenshot0000.eps,
width=0.3\textwidth}\hfill \epsfig{file=screenshot0001.eps,
width=0.3\textwidth} \hfill \epsfig{file=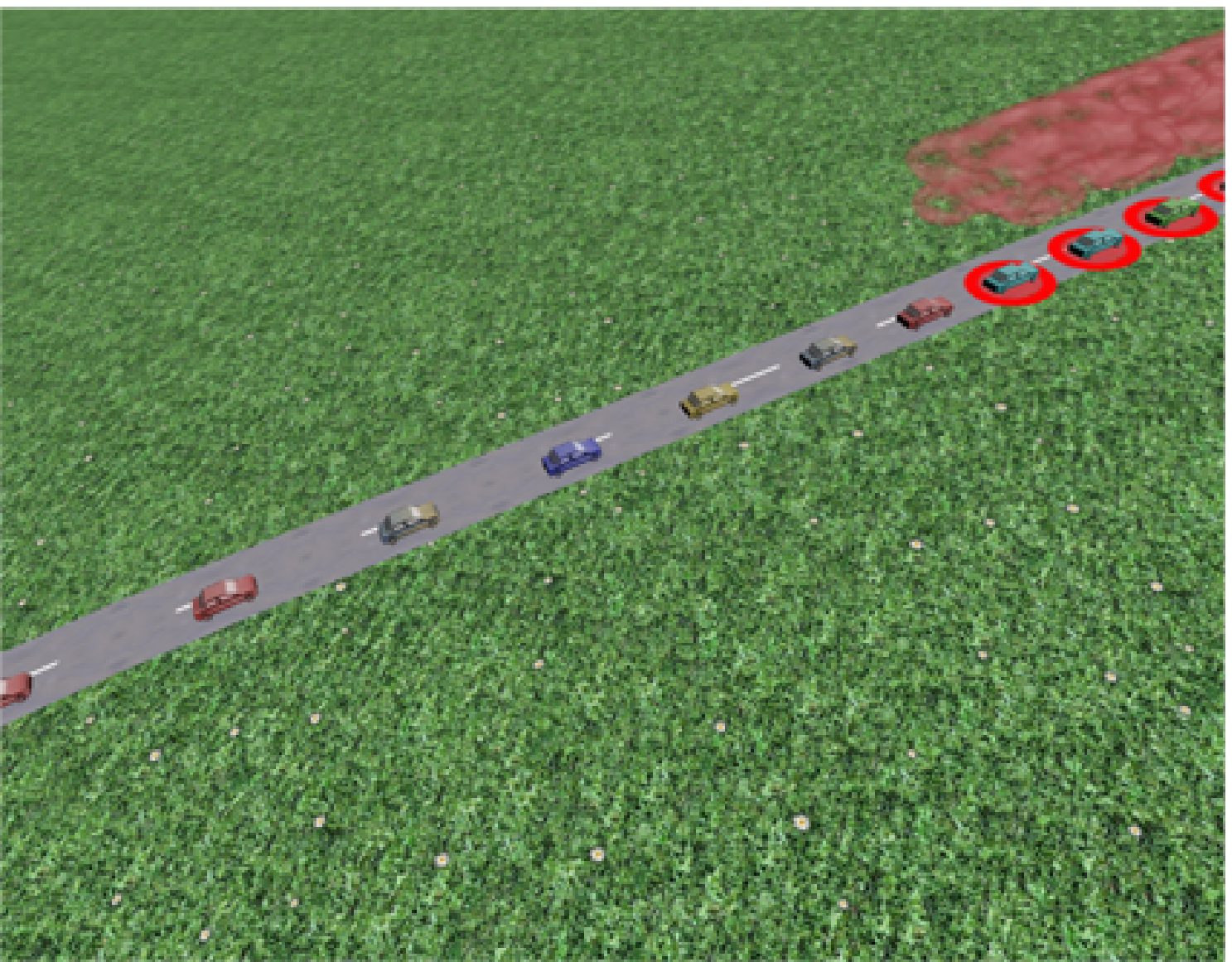, width =
0.3\textwidth} \caption{\label{isola}\small Screenshots from the
simulation of HDCs for recognizing a traffic jam. Left: front of
the traffic jam. Center: overview with front and end. Right: end
of the traffic jam. }
\end{figure}
%%%%%%%%%%%%%%%%%%%%%%%%%%%%%%%%%%%%%%%%%%%%%%%%%%%%%%%%%%%%%%%%%%%%%%

\iffalse
%%%%%%%%%%%%%%%%%%%%%%%%%%%%%%%%%%%%%%%%%%%%%%%%%%%%%%%%%%%%%%%%%%%%%%%
\begin{figure}[tp]
\centering \epsfig{file=screenshot0000.eps,
width=6cm}\\\vspace*{1cm} \epsfig{file=screenshot0001.eps,
width=6cm}
\ \\
\vspace*{1cm} \epsfig{file=screenshot0002.eps, width = 6cm}
\caption{\label{isola}\small Screenshots from the simulation of HDCs
for recognizing a traffic jam. Top: the front of the traffic jam.
Center: overview with front and end. Bottom: the end of the traffic
jam. }
\end{figure}
%%%%%%%%%%%%%%%%%%%%%%%%%%%%%%%%%%%%%%%%%%%%%%%%%%%%%%%%%%%%%%%%%%%%%%
\fi

\section{Distinguishing Types of Traffic Jams}

\subsection{Traffic Jams}
What exactly constitutes a traffic jam?
    According to \cite{k-mmtfi-97} a
    ``connected structure of vehicles, traveling at a velocity below a
    given threshold $v_{\mathrm{thresh}}$ will be called a jam if this
    structure contains at least one stopped vehicle.'' A good
estimate is $v_{\mathrm{thresh}} = \frac{v_{\mathrm{max}}}{2}$,
${v_{\mathrm{max}}}$ being the maximum velocity.

However, as every driver knows from his or her own experience, not
all traffic jams are alike; in particular, some appear to be an act
of nature (e.g., getting stuck in a blocked highway after an
accident), while others seem to have appeared out of thin air. When
trying to improve traffic flow and energy consumption of vehicles,
recognizing these distinctions is of crucial importance, as it
constitutes the prerequisite for developing successful strategies.

\subsection{Types of Traffic Jams}\label{subsec:types}
Another line of research investigates the characteristic properties
of congested traffic and tries to identify different congestion
patterns. According to current research \cite{sh-efctstitm-07},
there are five different basic types of traffic jams, which we will
present here in detail as we will use this categorization:

%As mentioned in Section~\ref{subsec:types} \cite{sh-efctstitm-07}
%distinguish five basic types of traffic jams: Pinned Localized
%Clusters (PLC), Oscillating Congested Traffic (OCT), Stop-and-Go
%Waves (SGW), Homogeneous Congested Traffic(HCT) and Moving Localized
%Clusters (MLC).

    \begin{itemize}
    \item Pinned Localized Cluster (PLC):\\
Neither the front or back of the congestion move upstream or
downstream, i.e., a traffic jam of fixed length is situated upstream
of a bottleneck. (If there is oscillation, this an oscillating
pinned localized
    cluster, OPLC).
A congestion of this type may arise spontaneously, or by upstream
traveling perturbation, which stops at the bottleneck. If the
density increases, this can change into a different type of
congestion, in particular into spatially extended congestion
patterns.

    \item Oscillating Congested Traffic (OCT)):\\
    This is a spatially extended congestion pattern, i.e., one that
grows in length, meaning that the upstream end moves in space. An
OCT may be caused by a perturbation or supercritical traffic
density; once the bottleneck has been removed, the congestion
gradually disappears. Characteristic for this type of congestion is
that a driver passing through it experiences ``more or less regular
oscillations of speed'', commonly known as stop-and-go traffic.
(However, this is different from the technical definition below.)
Frequency and amplitude of the oscillations are relatively stable,
and oscillations themselves move at about 15~km/h in the upstream
direction. OCT is surrounded by free traffic.

    \item Stop-and-Go Waves (SGW):\\
They are closely related to OCT; they have a large characteristic
amplitude, but no characteristic wave length. They have been
observed for a long time, the first description we are aware of
appears in \cite{ef-tft-58}. SGWs consist of a sequence of
congestions, separated by free traffic; individual congestions are
stable in length, so they are spatially confined.
    An average duration of a wave appears to lie between 4 and 20 minutes,
their typical speed is about 15 km/h in the upstream direction. SGW
arise either from small perturbations, or from PLCs; in most cases,
they turn into free-flowing traffic, but also into PLCs.

    \item Homogeneous Congested Traffic (HCT):\\
In HCT, speed is very low and relatively homogeneous over a larger
stretch of road. HCT tends to form a spatially extended congestion
pattern, with the downstream front fixed (usually at a bottleneck),
while the upstream front moves further upstream. When the bottleneck
is removed, the downstream front moves upstream at a typical speed
of 15~km/h.

    \item Moving Localized Cluster (MLC):\\
While SGW consist of several congestions, an individual one is
called a MLC. Both upstream and downstream front move at about
15~km/h upstream, with limited spatial dimensions. A MLC is caused
by a perturbation.
    \end{itemize}

\subsection{Distinguishing Congestion Types}
Any driver can decide that he is in a traffic jam by simply looking
out of the window. But what kind of congestion is it? This is a first
important step for actually changing the traffic situation by
allowing the right decisions to be made.

In Figure~\ref{jt} it is shown how this can be achieved within our
framework. In case a traffic jam emerges, this is observed and
Hovering Data Clouds are established at the back and front of the
jam. Considering the movement of these clouds over the time, as well
as the speed of the cars between these HDCs enables us to
distinguish the congestion types: A Pinned Localized Cluster is
characterized by a fixed upstream end, i.e., in case the back HDC
stays at a fixed location we may identify a PLC. In the event of a
Hovering Data Cloud at the back moving upstream, we consider the
velocity profile of the cars between the front and the back HDC:
oscillations of speed indicate an Oscillating Congested Traffic. To
identify Homogeneous Congested Traffic, it is now sufficient to look
at the distance between the front HDC and the back HDC. If this
distance increases over time, we may identify a spatially extended
traffic jam; with the other criteria: an HCT. For the final
distinction between Stop-and-Go Waves and a Moving Localized
Cluster, it suffices to check whether there are any nearby clusters
of similar amplitude. The lack of nearby clusters indicates a single
isolated traffic jam, here a MLC. Hence, keeping track of the
location of the HDCs, the velocity profile of the cars between and
messages on nearby clusters of congested traffic allows us to
differentiate the five congestion types.

%IPE-Abb.
%%%%%%%%%%%%%%%%%%%%%%%%%%%%%%%%%%%%%%%%%%%%%%%%%%%%%%%%%%%%%%%%%%%%%%%
%\begin{figure}[tp]
%\centering \epsfig{file=stautypen_diff.eps, width=.6\textwidth}
%\caption{\label{jt}Schematic distinction of traffic jam types.
%These considerations are carried out on the HDC level for each cluster. }
%\end{figure}
%%%%%%%%%%%%%%%%%%%%%%%%%%%%%%%%%%%%%%%%%%%%%%%%%%%%%%%%%%%%%%%%%%%%%%

\iffalse
%Inkspace-Abb.
%%%%%%%%%%%%%%%%%%%%%%%%%%%%%%%%%%%%%%%%%%%%%%%%%%%%%%%%%%%%%%%%%%%%%%%
\begin{figure}[tp]
\centering \epsfig{file=jamtypes_diff_2.eps, width=.6\textwidth}
\caption{\label{jt}\small Schematic distinction of traffic jam
types. These considerations are carried out on the HDC level for
each cluster. }
\end{figure}
%%%%%%%%%%%%%%%%%%%%%%%%%%%%%%%%%%%%%%%%%%%%%%%%%%%%%%%%%%%%%%%%%%%%%%
\fi

\iffalse
%IPE-Abb. quer
%%%%%%%%%%%%%%%%%%%%%%%%%%%%%%%%%%%%%%%%%%%%%%%%%%%%%%%%%%%%%%%%%%%%%%%
\begin{figure}[tp]
\centering \epsfig{file=jamtypes_diff_3.eps, width=.8\textwidth}
\caption{\label{jt}\small Schematic distinction of traffic jam
types. These considerations are carried out on the HDC level for
each cluster. }
\end{figure}
%%%%%%%%%%%%%%%%%%%%%%%%%%%%%%%%%%%%%%%%%%%%%%%%%%%%%%%%%%%%%%%%%%%%%%
\fi

%Inkscape-Abb. quer
%%%%%%%%%%%%%%%%%%%%%%%%%%%%%%%%%%%%%%%%%%%%%%%%%%%%%%%%%%%%%%%%%%%%%%%
\begin{figure}[tp]
\centering \epsfig{file=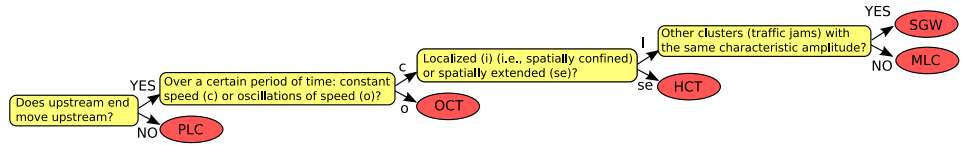, width=\textwidth}
\caption{\label{jt}\small Schematic distinction of traffic jam
types. These considerations are carried out on the HDC level for
each cluster. }
\end{figure}
%%%%%%%%%%%%%%%%%%%%%%%%%%%%%%%%%%%%%%%%%%%%%%%%%%%%%%%%%%%%%%%%%%%%%%

\section{Improving the Flow of Traffic}

The grand challenge for any kind of traffic research is to actually
improve the flow of traffic, in terms of travel time, 
energy consumption, or a combination of both. Over
the years, many different methods have been tried and even more have
been suggested, but quite clearly, most have been too crude to have
a real impact.

The methods for recognizing types of congestion are a first step
in this direction; this has an impact both on the microscopic level,
where making drivers aware of the type of traffic situation
has an impact on useful behavior, as well as the macroscopic
level, where routing is closely intertwined with forecasting,
which is dependent not just on the amount of existing congestion,
but also on its foreseeable development.
In particular, there are a number of actions that can be taken at different
levels:

\begin{itemize}
\item Identifying a pinned localized cluster comes with the clear recognition
of a cause. This makes it possible to pinpoint possible action
at the bottleneck, either by removing it, or by updating
medium-range traffic information if removal is impossible; in the latter
case, the identification of PLC means that
drivers can be sure they can proceed swiftly once they are
beyond the obstacle, which is not the case for stop-and-go waves.
\item While the cause for homogeneous congested traffic is similar
to that of a pinned localized cluster, the effect of removing the
bottleneck is different; furthermore, large-scale re-routing and
forecasting are absolutely vital for reducing further flow
into the jam, and thus prevent the congestion from causing
even larger problems.
\item The onset of oscillating traffic triggers a number
of different actions. On a small scale, driver behavior can be
influenced in order to improve flow; we have recently developed
methods that work surprisingly well for this purpose, results are
reported in a forthcoming paper~\cite{fth+-ifct-08}. On a
medium scale, increasing congestion may make it sensible to re-route
traffic; this requires dealing with the large-scale optimization
impact of flows over time (see \cite{skut} for a recent survey),
i.e., turning to more advanced optimization methods, which are
currently being studied in a separate context, see \cite{advest}.
\item Identifying stop-and-go waves is
particularly vital for
preventing crashes, as well as reducing unnecessary fuel consumption.
Moreover, distinguishing interior waves (where swift acceleration is
wasteful and even hazardous) as opposed to the front of the jam (where
outflow from the jam may be impeded by distrustful drivers) is critical for
the overall flow rate.
\item Identifying a moving localized cluster is particularly useful
for regulating outflow, and thus overall flow, by letting drivers realize
that there are no further oscillations ahead.
\end{itemize}

%%%%%%%%%%%%%%%%%%%%%%%%%%%%%%%%%%%%%%%%%%%%%%%%%%%%%%%%%%%%%%%%%%%%%%%%%%%%%%%%%%%

\begin{figure*}[!t]
        \begin{center}
                \mbox
                {
            \subfigure[]
            { \includegraphics[width=0.45\columnwidth] {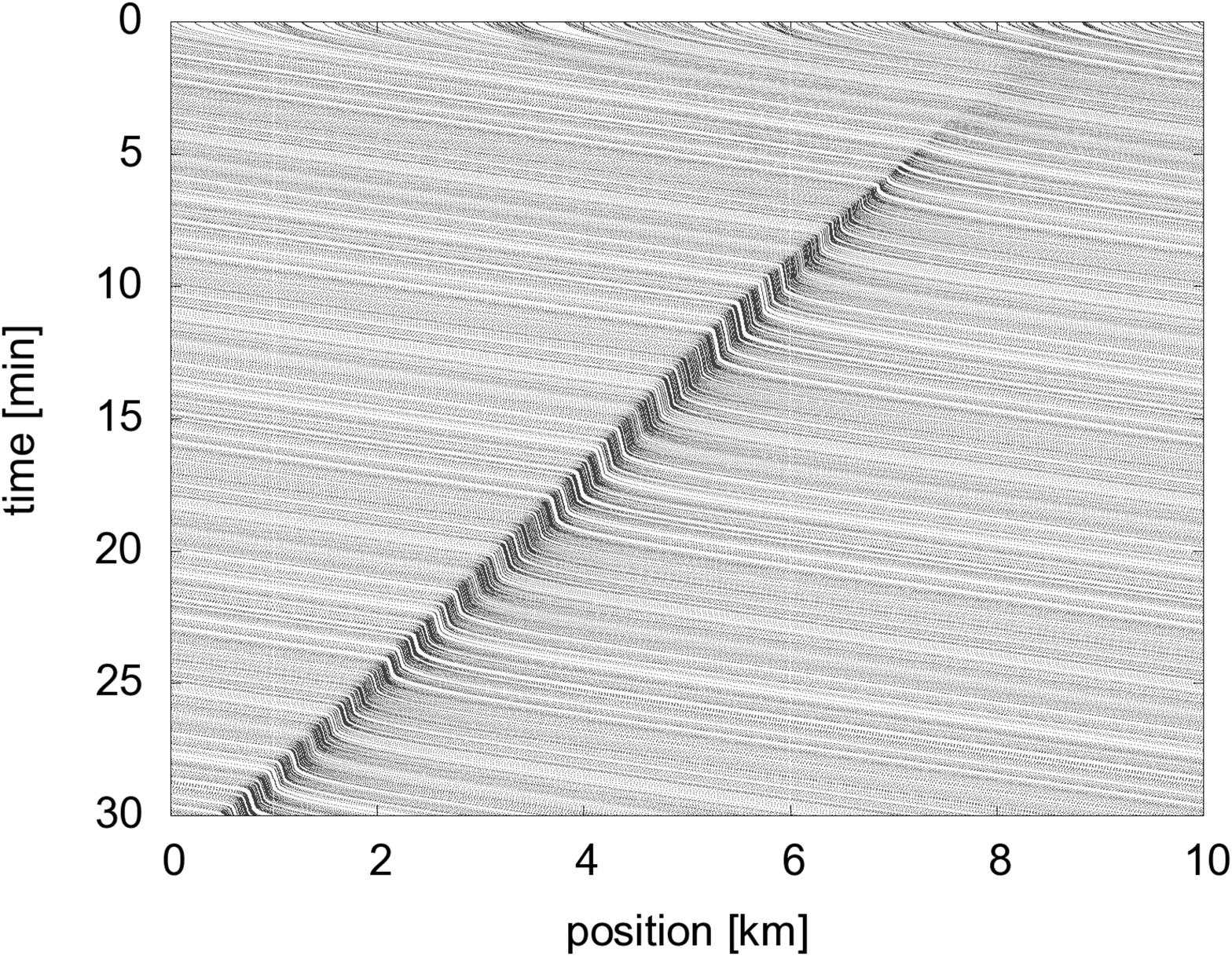} \label{fig:carflow105}}
           \quad
            \subfigure[]
            { \includegraphics[width=0.45\columnwidth] {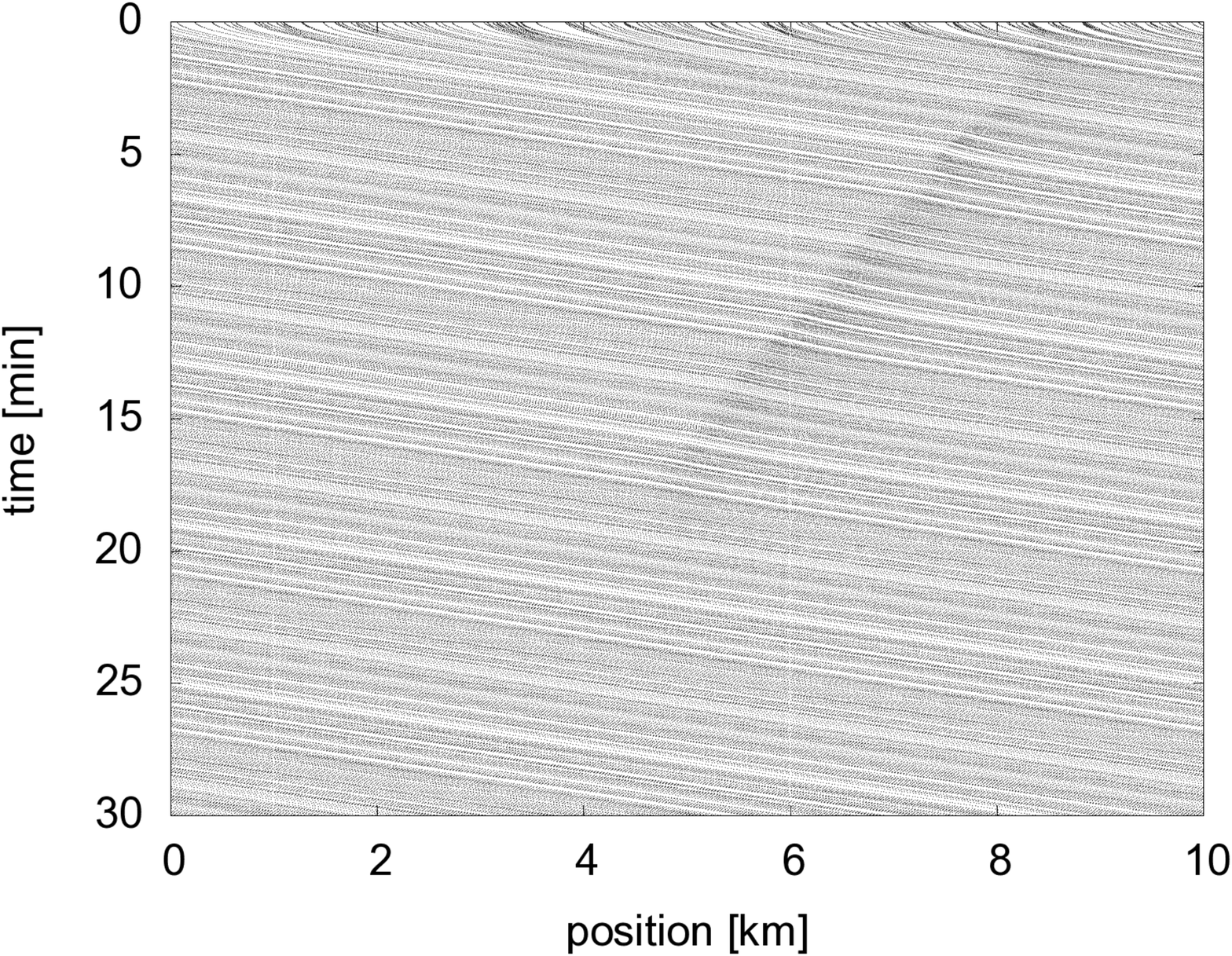} \label{fig:carflow100}}
                }

          \caption{\small Illustration of vehicle flow over time (traffic density $\rho=30$\,veh/km). \figref{carflow105} shows a traffic jam that emerges from local disturbances, with a speed limit of 105\,km/h. Limiting the maximum speed by just 5\,km/h to 100\,km/h will avoid the traffic jam, as depicted in \figref{carflow100}.}
\label{fig:carflow}

        \end{center}
%\vspace*{-0.6cm}
\end{figure*}

%%%%%%%%%%%%%%%%%%%%%%%%%%%%%%%%%%%%%%%%%%%%%%%%%%%%%%%%%%%%%%%%%%%%%%%%%%%%%%%%%%%

Another way to improvement is illustrated in the two parts of \figref{carflow}.
In both diagrams each line depicts the trajectory of a vehicle on the road: the
faster the travel speed of the vehicle, the lower the incline of the line. As
vehicles get slower, the gradient of the lines increases with time.
Due to individual behavior of drivers, a traffic jam occurs in
\figref{carflow105} at an unpredictable time and position.
\figref{carflow100} shows a very similar scenario, in which the
above jam does not occur. We have the same amount of cars and types,
but reduce the global speed limit from 105\,km/h to 100\,km/h. The
simulation confirms that small changes in the setup (like a slight
reduction in the speed of cars) result in very different and
unpredictable traffic structures.

Previous work on these issues suggests implementing a dynamic global
speed limit to reduce the occurrence of traffic jams (see
\cite{treiber-2001-49}). Although the speed limit is globally
assigned, and thereby a very coarse-grained measure, it suffices in
our scenario to avoid the traffic jam.

A more sophisticated approach was described in
\cite{ktsfh-jaacciiotd-05}: the authors propose a system called
Adaptive Cruise Control (ACC) that
    ``automatically accelerates or decelerates a vehicle to
    maintain a selected gap, to reach a desired velocity, or to
prevent a rear-end collision.'' It is based on car-following methods
(i.e., each vehicle adapts its behavior to distance and velocity of
the car preceding it.) As the authors demonstrated in simulations,
automated driving strategies help to improve capacity and stability
of traffic flow, even if only a limited fraction of vehicles are
equipped with ACC. However,
ACC does not make use of higher-level communication and coordination
capabilities, which is what we are aiming for.

We foresee a huge potential for quick and local reactions in a
future traffic information system with only a small impact for the
single car. In general, it is a difficult task to evaluate the
critical thresholds at which traffic structures evolve (as is the
case for the traffic jam in \figref{carflow}), in particular in
dynamically changing traffic densities, as occurring in real life.
We believe that our framework allows going a step even beyond ACC,
by actually using the communication capabilities for self-organized
and coordinated behavior of a whole group of vehicles, instead of
just setting global or local parameters.

\section {Organic Information Complexes}
\label{sec:autonomos_oic}
%%%%%%%%%%%%%%%%%%%%%%%%%%%%%%%%%%%%%%%%%%%%%%%%%%%%%%%%%%%%%%%%%%%%%%%%%%%
%%%%%%%%%%%%%%%%%%%%%%%%%%%%%%%%%%%%%%%%%%%%%%%%%%%%%%%%%%%%%%%%%%%%%%%%%%%

\begin{figure*}[!t]
        \begin{center}
                \includegraphics[scale=0.5] {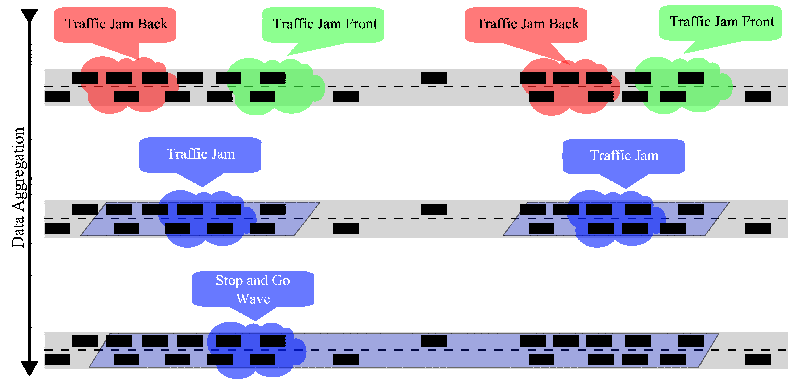}
                \caption{\small Data aggregation by an Organic Information Complex to recognize a traffic jam.}
                \label{fig:data_aggregation_jam}
        \end{center}
\end{figure*}
We conclude by sketching the next step beyond Hovering Data Clouds:
using them to form larger-scale structures. We call these resulting
structures {\em Organic Information Complexes} (OICs). While
some of the resulting ideas are closely related to ideas
described above, further aspects of OICs are subject to future work.

As described above, taking appropriate action may require finding means to
describe higher-level, more complex structures, e.g., the individual
stop-and-go waves in a SGW congestion.
The data gathered by HDCs is raw data, e.g., describing the end of a
traffic jam. We need to combine several HDCs to obtain higher-level
information. As shown in \figref{data_aggregation_jam}, HDCs
resulting from the locally identified phenomena, e.g., {\em back of
traffic jam } and {\em front of traffic jam} send out data units.
When those data units accumulate, higher-level information can be
built that describes the congestion from a higher perspective. When
{\em front-} and {\em back-}HDCs gather a higher-level construct, an
{\em Organic Information Complex} (OIC) arises. In our example the
two {\em traffic jam}-OICs form a {\em stop and go wave}-OIC. This
merging and aggregation of simple data into higher-value data is the
basic idea for building up {\em Organic Information Complexes}
(OICs).

\begin{figure*}[!t]
        \begin{center}
                \includegraphics[scale=0.5] {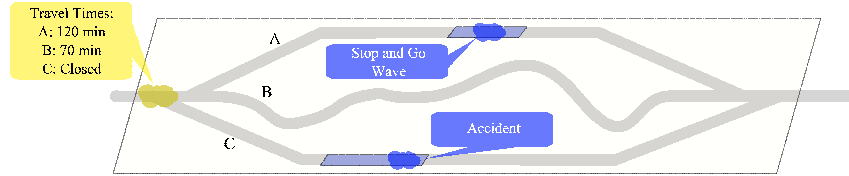}
                \caption{\small An Organic Information Complex on a road network aggregates possible routes.}
                \label{fig:data_aggregation_road_network}
        \end{center}
\end{figure*}

A more complex scenario is depicted by
\figref{data_aggregation_road_network}, where OICs have discovered
congestions on roads {\em A} and {\em C} and react by sending this
information towards incoming vehicles. When this information reaches
intersections on the road network that can be utilized for bypassing
the traffic jams, the different pieces of information can be turned
into a journey-time prediction for the different routes.

Due to this in-network data aggregation, the overall system remains
scalable. Note that HDCs and OICs are not necessarily bound to
fixed locations, nor to specific network nodes. They arise in a
self-organized manner, wherever the matching conditions occur. In
almost the same manner, aggregation is not restricted to fixed
locations, but happens as a result of appropriate concurring data
units.

    \section{Outlook}

We have described a number of aspects of using wireless
communication as the basis for self-organizing methods for
participants in traffic. As demonstrated, this work extends from
local protocols for communication all the way to high-level
organization; methodically, it involves practical traffic models and
simulations, as well as theoretical distributed algorithms.
At this point, we have laid the basis for communication and coordination,
and we have dealt with data dissemination; we were also able to demonstrate
lower-level self-organization and self-recognition by making use of
our concept of Hovering Data Clouds. First steps towards self-modification
are on their way, and higher-level aspects are to follow.

Clearly, this is work in progress. We are optimistic that
AutoNomos will continue to produce interesting results; even just
realizing our ideas at an intermediate level may have far-reaching
benefits for the way we act and interact in traffic.

    %%%%%%%%%%%%%%%%%%%%%%%%%%%%%%%%%%%%%%%%%%%%%%%%%%%%%%%%%%%%%
    % Bibliography
    %%%%%%%%%%%%%%%%%%%%%%%%%%%%%%%%%%%%%%%%%%%%%%%%%%%%%%%%%%%%%
    \bibliographystyle{acmtrans}
    \bibliography{ref_taas,vtc,isola}

\newpage

\section*{Appendix A}

\definecolor{light}{gray}{0.6}

\begin{algorithm}
\footnotesize \underline{\textbf{Algorithm 1:}} \textbf{HDCs for
traffic jams}\\ \vspace*{0.3cm}
%hier ggf. wieder settimer($clock$ , smData);\\
\textbf{initialization:}\\
 $i=0;$ \textit{etc.} \; {\em buffer} = $\emptyset$\;

 {\em settimer(clock, smData)};

\SetKwBlock{oTsD}{onTimer(smData)}{} \SetKwComment{kommentar}{//}{}
\SetKwBlock{oTD}{onTimer(Data)}{}
\SetKwBlock{LBr}{LBrecv(\textit{m})}{} \SetKwBlock{leerB}{}{}
\SetKwIf{BlIf}{BlEls}{\If{}{}}{then}{else}{endif}

\SetNoline
%%%%%%%%%%%%%%%%%%%%%%%%%%%%%%%%%%%%%%%%%%%%%%%%%%%%%%%%%%%%%%%%%%%%%%
%onTimer(smData)
%%%%%%%%%%%%%%%%%%%%%%%%%%%%%%%%%%%%%%%%%%%%%%%%%%%%%%%%%%%%%%%%%%%%%%
\oTsD{ $q$\_before = $Q\_v$\; $p$\_before = $p\_v$\; {\em back} =
$\infty$\;   {\em front} = 0\; $P\_v$ = 0; $Q\_v$ = 0\; $p\_v$ = 0;
$q\_v$ = 0\; $p$ = 0; $q$ = 0\;

{\em LBcast-CR($\langle$smdata, ID, loc, v$\rangle$)}\mbox{;} \CommentSty{  //broadcast in CR}\\
\textit{settimer(next-multiple($t_{\mathrm{data}}$), Data)}\mbox{;}
\CommentSty{  //as Data follows smData}\\ $s = 0$;}
%%%%%%%%%%%%%%%%%%%%%%%%%%%%%%%%%%%%%%%%%%%%%%%%%%%%%%%%%%%%%%%%%%%%%%
%onTimer(Data)
%%%%%%%%%%%%%%%%%%%%%%%%%%%%%%%%%%%%%%%%%%%%%%%%%%%%%%%%%%%%%%%%%%%%%%
\oTD{\textit{t} = \textit{clock}\; \textit{congestion\_counter =
0}\; \SetVline \eIf{\SetNoline(congestion\_participant = 1) $\vee$
\\\hspace*{0.5cm}((p = 1) $\wedge$ (congestion\_participant\_before = 1)) $\vee$
\\\hspace*{0.5cm}((q = 1) $\wedge$ (congestion\_participant\_before = 1))}{\SetVline
\textit{congestion\_participant\_before = 1}; }{
\textit{congestion\_participant = 0}\; $p$\_before = 0\;$q$\_before
= 0\;} \If{(\textit{status}$_\mathrm{back}$ =
\textit{active}$_\mathrm{back}$) $\wedge$ (congestion\_participant =
0)}{\textit{status}$_\mathrm{back}$ = idle\;ahead = 0\;}
\If{(\textit{status}$_\mathrm{front}$ =
\textit{active}$_\mathrm{front}$) $\wedge$ (congestion\_participant
= 0)}{\textit{status}$_\mathrm{front}$ = idle\;}
%\BlankLine
%}
%\end{algorithm}
%
%\begin{algorithm}
%\SetKwBlock{LBr}{LBrecv(\textit{m})}{} \SetKwBlock{leerB}{}{}
%\SetKwIf{BlIf}{BlEls}{\If{}{}}{then}{else}{endif}
%\small
%{
\If{(congestion\_participant = 0 $\wedge$ \textit{v} $\leq$ CVmax)
$\vee$
\\\hspace*{0.5cm}(\textit{status}$_\mathrm{back}$ = \textit{active}$_\mathrm{back}$) $\vee$
\\\hspace*{0.5cm}(\textit{status}$_\mathrm{front}$ =
\textit{active}$_\mathrm{front}$)}{\textit{LBcast($\langle$\textit{data},
ID, $loc, v, p,
q \rangle$)}\mbox{;} \CommentSty{ //broadcast in R}\\
\textit{settimer(next-multiple($t_{\mathrm{smdata}}$),
smData)}\mbox{;} \CommentSty{//as
smData follows Data}\\
\textit{congestion\_participant\_before = congestion\_participant}\;
\textit{congestion\_participant = 0}\;}
\gElsIf{(congestion\_participant = 1) $\vee$
\\\hspace*{1cm}(congestion\_participant = 0 $\wedge$ \textit{v} $>$
CVmax)}{\textit{congestion\_participant\_before =
congestion\_participant\; congestion\_participant = 0\;
settimer(next-multiple(\textit{t}$_{\mathrm{smdata}}$),
smData)}\mbox{;} \CommentSty{//as smData follows Data}\\} \textit{i}
= 0\; }
%\BlankLine
%%%%%%%%%%%%%%%%%%%%%%%%%%%%%%%%%%%%%%%%%%%%%%%%%%%%%%%%%%%%%%%%%%%%%%
%onTimer(LBrecv(m))
%%%%%%%%%%%%%%%%%%%%%%%%%%%%%%%%%%%%%%%%%%%%%%%%%%%%%%%%%%%%%%%%%%%%%%
\LBr{\textit{buffer = buffer} $\cup\,\, \langle$ \textit{m,
clock}$\rangle$\; \textit{settimer(\textit{clock + d},
NewMessage)}\;}
\end{algorithm}

\begin{algorithm}
\footnotesize \SetKwBlock{oTC}{Congestion}{}
\SetKwBlock{oTI}{onTimer(Information)}{}
\SetKwBlock{oTCA}{CongestionAhead}{}
\SetKwBlock{oTAI}{onTimer(AheadInfo)}{}
\SetKwBlock{LBr}{LBrecv(\textit{m})}{}

\SetNoline %\LBr{\textit{buffer = buffer} $\cup\,\, \langle$
%\textit{m, clock}$\rangle$; \textit{settimer(\textit{clock + d},
%NewMessage)}\;}
%%%%%%%%%%%%%%%%%%%%%%%%%%%%%%%%%%%%%%%%%%%%%%%%%%%%%%%%%%%%%%%%%%%%%%
%Congestion
%%%%%%%%%%%%%%%%%%%%%%%%%%%%%%%%%%%%%%%%%%%%%%%%%%%%%%%%%%%%%%%%%%%%%%
\oTC{\SetVline \If{\textit{status}$_\mathrm{back}$ =
\textit{active}$_\mathrm{back}$}{
\If{(\textit{location}$_\mathrm{back}$ $<$ back) $\vee$
(\textit{location}$_\mathrm{back}$ $>$
back)}{\textit{location}$_\mathrm{back}$ = back\;}
\textit{settimer(next-multiple($t_{\mathrm{information}}$),
Information)}\;} \If{\textit{status}$_\mathrm{back}$ =
\textit{joining}$_\mathrm{back}$}{\textit{status}$_\mathrm{back}$ =
\textit{active}$_\mathrm{back}$\;
\textit{settimer(next-multiple($t_{\mathrm{information}}$),
Information)}\; \textit{location}$_\mathrm{back}$ = back\;}
\If{\textit{status}$_\mathrm{back}$ = \textit{idle}}{
\eIf{$q$\_before = 0}{\textit{location}$_\mathrm{back}$ = back\;
{\textit{state} = $v_{0_\mathrm{back}}$}\;
\textit{status}$_\mathrm{back}$ = \textit{active}$_\mathrm{back}$\;
\textit{settimer(next-multiple($t_{\mathrm{information}}$),
Information)}\;}{\textit{status}$_\mathrm{back}$ =
\textit{joining}$_\mathrm{back}$\mbox{;}\CommentSty{  //status to
collect all
dates}}}} %\BlankLine
\SetNoline
%%%%%%%%%%%%%%%%%%%%%%%%%%%%%%%%%%%%%%%%%%%%%%%%%%%%%%%%%%%%%%%%%%%%%%
%onTimer(onTimer(Information))
%%%%%%%%%%%%%%%%%%%%%%%%%%%%%%%%%%%%%%%%%%%%%%%%%%%%%%%%%%%%%%%%%%%%%%
\oTI{\SetVline\If{\textit{status}$_\mathrm{back}$ =
\textit{active}$_\mathrm{back}$}{
\textit{LBcast($\langle$\textit{Congestion},
{location}$_\mathrm{back}$, HDCfront\_location, v} $\rangle$)\;
settimer(next-multiple($t_{\mathrm{information}}$),
Information)\;%\mbox{;} \CommentSty{//broadcast the information on the traffic jam}\\
\If{$|$\textit{loc-location}$_\mathrm{back}| >
r_{\mathrm{HDC}}$}{\textit{status}$_\mathrm{back}$ =
\textit{idle}\mbox{;} \CommentSty{//If the distance is too big, the processor// becomes idle. So, on the next call of \textit{Information} no further //timer is set}\\
{\em ahead = 0}\;}}}
%\BlankLine
\SetNoline
%%%%%%%%%%%%%%%%%%%%%%%%%%%%%%%%%%%%%%%%%%%%%%%%%%%%%%%%%%%%%%%%%%%%%%
%onTimer(CongestionAhead)
%%%%%%%%%%%%%%%%%%%%%%%%%%%%%%%%%%%%%%%%%%%%%%%%%%%%%%%%%%%%%%%%%%%%%%
\oTCA{\SetVline \If{\textit{status}$_\mathrm{front}$ =
\textit{active}$_\mathrm{front}$}{
\If{(\textit{location}$_\mathrm{front} <$ front) $\vee$
(\textit{location}$_\mathrm{front} >$
front)}{\textit{location}$_\mathrm{front}$ = front\;}
\textit{settimer(next-multiple($t_{\mathrm{aheadInfo}}$),
AheadInfo)}\;} \If{\textit{status}$_\mathrm{front}$ =
\textit{joining}$_\mathrm{front}$}{\textit{status}$_\mathrm{front}$
= \textit{active}$_\mathrm{front}$\;
\textit{settimer(next-multiple($t_{\mathrm{aheadInfo}}$),
AheadInfo)}\; \textit{location}$_\mathit{front}$ = \textit{front}\;}
\If{\textit{status}$_\mathit{front}$ = \textit{idle}}{
\eIf{$p$\_before = 0}{\textit{location}$_\mathrm{front}$ = {em
front}\; {\textit{state} = $v_{0_\mathrm{front}}$}\;
\textit{status}$_\mathrm{front}$ =
\textit{active}$_\mathrm{front}$\;
\textit{settimer(next-multiple($t_{\mathrm{aheadInfo}}$),
AheadInfo)}\;}{\textit{status}$_\mathrm{front}$ =
\textit{joining}$_\mathrm{front}$;}}}

%\BlankLine
\SetNoline
%%%%%%%%%%%%%%%%%%%%%%%%%%%%%%%%%%%%%%%%%%%%%%%%%%%%%%%%%%%%%%%%%%%%%%
%onTimer(onTimer(AheadInfo))
%%%%%%%%%%%%%%%%%%%%%%%%%%%%%%%%%%%%%%%%%%%%%%%%%%%%%%%%%%%%%%%%%%%%%%
\oTAI{\SetVline \If{\textit{status}$_\mathrm{front}$ =
\textit{active}$_\mathrm{front}$}{\em
LBcast($\langle$\textit{hdcdistance},
\textit{location}$_\mathrm{front} \rangle$);}
\If{$|$\textit{loc-location}$_\mathrm{front}| >
r_{\mathrm{HDC}}$}{\textit{status}$_\mathrm{front}$ =
\textit{idle}\;}}

\end{algorithm}

\begin{algorithm}%\caption{\label{alg} HDCs for traffic jams}
\footnotesize \SetKwBlock{oTNM}{onTimer(NewMessage)}{}
%\SetKwIf{CIf}{CElse}{if}{then{\CommentSty{//checking critical values
%}}{else}{endif}

\SetNoline
%%%%%%%%%%%%%%%%%%%%%%%%%%%%%%%%%%%%%%%%%%%%%%%%%%%%%%%%%%%%%%%%%%%%%%
%onTimer(onTimer(NewMessage))
%%%%%%%%%%%%%%%%%%%%%%%%%%%%%%%%%%%%%%%%%%%%%%%%%%%%%%%%%%%%%%%%%%%%%%
\oTNM{\SetVline let \textit{m} =
\textit{min}(\textit{m}:$\langle$\textit{ m, t} $\rangle \in$
buffer, \textit{t} = \textit{clock - d})\; \If{\textit{m} =
$\langle$\textit{data}, ident, \textit{l, g, p-value,
q-value}$\rangle$)}{\textnormal{\CommentSty{//checking critical
values}}\\\textit{env}$_i$ = $\langle$\textit{ident, l,
g}$\rangle$\; \If{($|$ \textit{loc - l} $| <$ CR) $\wedge$
(\textit{v} $<$ CV)}{back = \textit{min}\{back, \textit{loc, l}\}\;
\lIf{back==\textit{loc}}{back\_id = ID\;}
\lIf{back==\textit{l}}{back\_id = \textit{env}$_i$.\textit{ident}\;}
\textit{P}$\_v$ = \textit{p-value} of back\_id; \textit{Q}$\_v$ =
\textit{q-value} of back\_id\; front = \textit{max}\{front,
\textit{loc, l} \}\; \lIf{front==\textit{loc}}{front\_id = ID\;}
\lIf{front==\textit{l}}{front\_id =
\textit{env}$_i$.\textit{ident}\;} \textit{p}$\_v$ =
\textit{p-value} of front\_id; \textit{q}$\_v$ = \textit{q-value} of
front\_id\; congestion\_counter++; s = 1\; congestion\_participant =
1\;} \textnormal{\CommentSty{//comparison: transmitting processor -
already received messages}}\\ \For{$j,1,i-1$}{\If{($|$ \textit{l} -
\textit{env}$_j$.\textit{l} $| <$ CR) $\wedge$ (\textit{g} $<$
CV)}{back = \textit{min}\{back, \textit{l},
\textit{env}$_j$.\textit{l}\}\;
\lIf{back==\textit{env}$_j$.\textit{l}}{back\_id =
env}$_l$.\textit{ident}\; \lIf{back==\textit{l}}{back\_id =
\textit{env}$_i$.\textit{ident}\;}
 \textit{P}$\_v$ = \textit{p-value} of
back\_id; \textit{Q}$\_v$ = \textit{q-value} of back\_id\; front =
\textit{max}\{front, \textit{l}, \textit{env}$_j$.\textit{l} \}\;
\lIf{front==\textit{env}$_j$.\textit{l}}{front\_id =
\textit{env}$_j$.\textit{ident}\;} \lIf{front==\textit{l}}{front\_id
= \textit{env}$_i$.\textit{ident}\;} \textit{p}$\_v$ =
\textit{p-value} of front\_id; \textit{q}$\_v$ = \textit{q-value} of
front\_id\; \If{s = 0}{congestion\_counter++\; s = 1\;}}}}
%\BlankLine
\If{(congestion\_counter $> 0$) $\wedge$ ($|$back - \textit{loc}$| <
r_{\mathrm{HDC}}$) $\wedge$ (\textit{P}$\_v$ =
0)\\\hspace*{0.5cm}$\wedge$ (clock $\geq $\textit{t +
t}$_{\mathrm{ab}}$ + \textit{d})}{invoke Congestion\;
congestion\_participant = 1\;} \If{(congestion\_counter $> 0$)
$\wedge$ ($|$front - \textit{loc}$| < r_{\mathrm{HDC}}$) $\wedge$
(\textit{q}$\_v$ = 0)\\\hspace*{0.5cm} $\wedge$ (\textit{clock}
$\geq$ \textit{t + t}$_{\mathrm{ab}}$ + \textit{d})}{invoke
CongestionAhead\; congestion\_participant = 1\;} \lIf{ahead =
0}{HDCfront\_location = front; \textit{i}++\;}
%\BlankLine
\If{\textit{m} = $\langle$\textit{Congestion}, \textit{l},
location\_hdc\_front, \textit{g}$\rangle$}{\lIf{\textit{l} $>$
\textit{loc}}{LBcast($\langle$\textit{Congestion}, \textit{l},
\textit{location\_hdc\_front},
\textit{g}$\rangle$)\;}\lIf{$|$\textit{l - loc}$| <
r_{HDC}$}{\textit{status}$_\mathrm{back}$ =
\textit{joining}$_\mathrm{back}$;}}
%\BlankLine
\If{\textit{m} = $\langle$\textit{hdcdistance},
\textit{L}$_\mathrm{front}
\rangle$}{\If{\textit{status}$_\mathrm{back}$ =
\textit{active}$_\mathrm{back}$}{HDCfront\_location =
\textit{L}$_\mathrm{front}$\mbox{;} ahead =
1\;}\gElsIf{congestion\_participant =
1}{LBcast($\langle$\textit{hdcdistance}, \textit{L}$_\mathrm{front}
\rangle$);}}
%\BlankLine
\If{\textit{m} = $\langle$\textit{smData}, ident, \textit{l,
g}$\rangle$}{\lIf{\textit{l} $<$ \textit{loc}}{\textit{p} =
1\mbox{;}\CommentSty{  //right neighbor}\\} \lIf{\textit{l} $>$
\textit{loc}}{\textit{q} = 1\mbox{;}\CommentSty{  //left neighbor}}
}}
\end{algorithm}

    \end{document}